\documentclass[%
 reprint,
 superscriptaddress,
 amsmath,amssymb,
 aps,
prstab,
]{revtex4-2}

\usepackage{graphicx}
\usepackage{dcolumn}
\usepackage{bm}
\usepackage{hyperref}
\hypersetup{
    colorlinks=true,
    linkcolor=blue,
    filecolor=blue,
    urlcolor=blue,
    citecolor=blue,
}
\usepackage{siunitx}

\begin{document}

\title{5D tomographic phase-space reconstruction of particle bunches}


\author{S. Jaster-Merz}
\email[]{sonja.jaster-merz@desy.de}
\affiliation{Deutsches Elektronen-Synchrotron DESY, Germany}
\affiliation{Department of Physics Universität Hamburg, Germany}

\author{R. W. Assmann}
\affiliation{Deutsches Elektronen-Synchrotron DESY, Germany}
\affiliation{Laboratori Nazionali di Frascati, Italy}
\author{R. Brinkmann}
\author{F. Burkart}
\affiliation{Deutsches Elektronen-Synchrotron DESY, Germany}
\author{W. Hillert}
\affiliation{Department of Physics Universität Hamburg, Germany}
\author{M. Stanitzki}
\author{T. Vinatier}
\affiliation{Deutsches Elektronen-Synchrotron DESY, Germany}


\date{\today}

\begin{abstract}
    We propose a new beam diagnostics method to reconstruct the phase space of charged particle bunches in 5 dimensions, which consist of the horizontal and vertical positions and divergences as well as the time axis.
    This is achieved by combining a quadrupole-based transverse phase-space tomography with the adjustable streaking angle of a polarizable X-band transverse deflection structure (PolariX TDS).
    We demonstrate with detailed simulations that the method is able to reconstruct various complex phase-space distributions and that the quality of the reconstruction depends on the number of input projections. 
    This method allows for the identification and visualization of previously unnoticed detailed features in the phase-space distribution, and can thereby be used as a tool towards improving the performance of particle accelerators, or performing more accurate simulation studies. 
\end{abstract}


\maketitle

\section{INTRODUCTION}
\begin{figure*}
    \centering
    \includegraphics*[width=.9\textwidth]{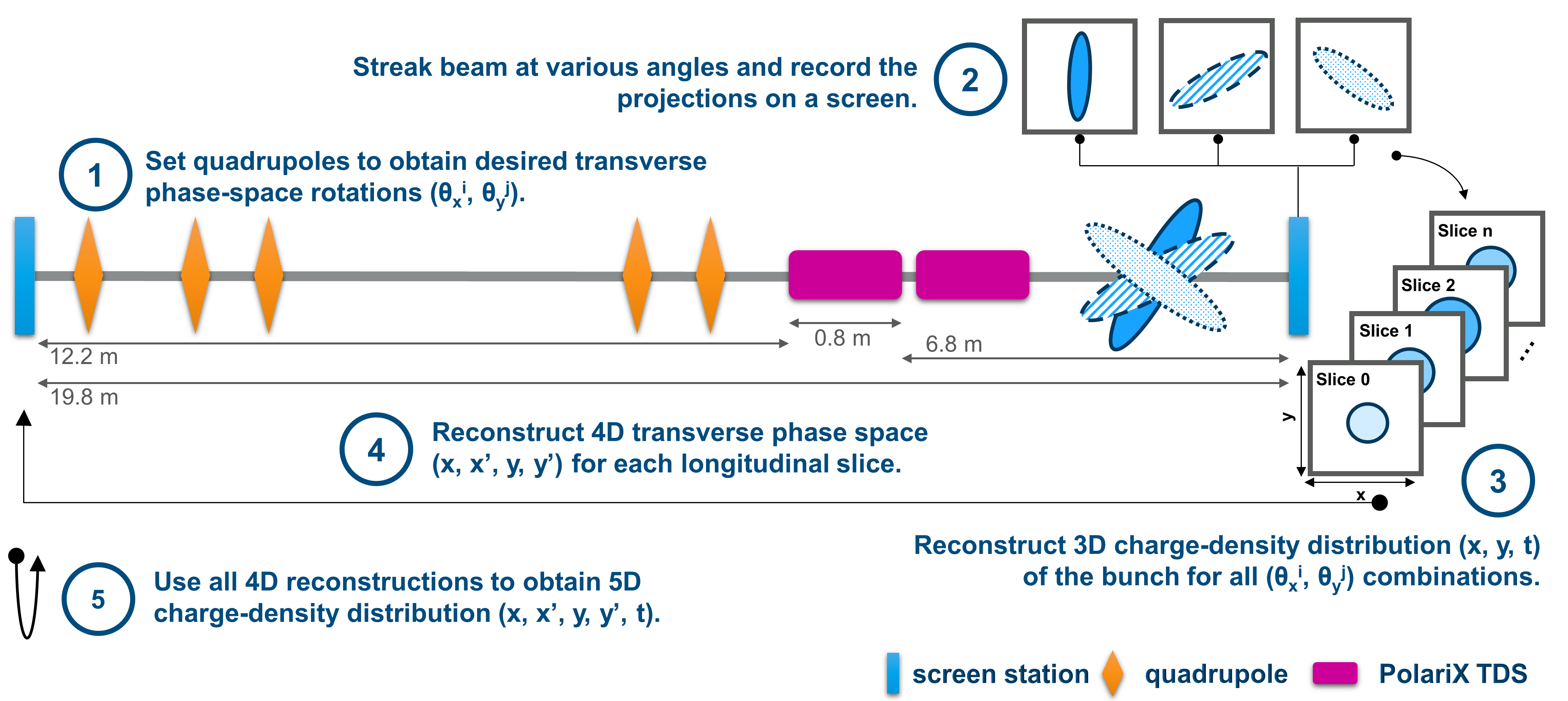}
    \caption{Sketch of the ARES beamline layout used for the 5D phase-space tomography simulation studies and working principle of the method. The gap between the third and fourth quadrupole is occupied by a bunch compressor which is not used for the presented study. Two PolariX TDSs are present in the ARES beamline. As mentioned in the text, only the first TDS is used in the presented simulation study.}
    \label{fig:beamline}
\end{figure*}
Detailed knowledge of the particle beam properties is required for simulating, optimizing and operating high-quality particle accelerators. 
Ideally, the full phase-space distribution should be known, however, only lower dimensional projections and statistical beam parameters are typically measured. 
While the full phase-space distribution is available from start-to-end simulation studies, such particle distributions do not always accurately represent the actual beam. This is due to, for example, an incomplete knowledge of the beamline and its instabilities like drifts and jitters or an incomplete knowledge of the initial properties of the beam.

At the same time, imperfections in the beamline or collective effects can impact the phase-space distribution and lead to correlations between the planes, or complex, non-Gaussian phase-space distributions that are not well described by lower dimensional projections or the statistical beam parameters.
For example, space-charge forces \cite{Ferrario2014} or coherent-synchrotron radiation \cite{dohlus2006bunch} can lead to correlations between the longitudinal and transverse planes. Rotated beamline elements or stray magnetic fields on the photocathode and in the gun region can lead to 2-dimensional (2D) emittance growth \cite{PhysRevAccelBeams.22.072805, PhysRevAccelBeams.21.010101}. 
For ultra-short bunches in the fs-range the energy gain deviations induced by transverse offsets in a radio frequency (RF) gun and the travel time difference induced by transverse offsets in transverse focusing elements also affect the phase space distribution \cite{PhysRevSTAB.9.084201}.

To characterize such effects in detail, diagnostic methods to reconstruct high-dimensional phase-space distributions are required.
For a low-energy \SI{2.5}{MeV} $\mathrm{H}^{-}$ beam, a measurement of the 6D charge density has been performed \cite{PhysRevLett.121.064804} in a dedicated beamline with slit masks. 
As an alternative, tomographic methods can be used which are readily applicable in most existing beamlines to access phase-space distributions in 2D with micrometer transverse resolution \cite{MCKEE1995264, PhysRevSTAB.9.112801, PhysRevSTAB.6.122801, PhysRevAccelBeams.24.022802}.
Methods exist to measure higher-dimensional distributions, such as the time-resolved transverse phase-space distribution \cite{PhysRevSTAB.12.050704}, the full 4D transverse phase space \cite{HOCK20138, PhysRevAccelBeams.23.032804, guo:ibic21-tupp15, jaster-merz:ipac21-mopab302}, as well as the 3D charge density distribution\cite{Marx_2017, Marx2019, Marx:427780, Marchetti2021}.
Furthermore, efforts to obtain information on the longitudinal phase space \cite{PhysRevLett.121.044801} and projections of the 6D phase space \cite{Scheinker2022} are made using machine learning techniques.

In recent years, a collaboration between CERN, DESY, and PSI has successfully developed and tested a novel polarizable X-band transverse deflection structure (PolariX TDS) \cite{PhysRevAccelBeams.23.112001, Marchetti2021, Grudiev:2158484}. In addition to the conventional TDS features, this structure is capable of streaking bunches along any transverse direction, opening up new opportunities for more detailed diagnostic methods, such as a recently experimentally demonstrated 3D charge-density reconstruction \cite{Marchetti2021}. Several PolariX TDSs have been installed at DESY and PSI, including two structures at the ARES linear accelerator \cite{burkart:linac2022-thpojo01, Marchetti_2020, instruments5030028, DORDA2018239}, a facility dedicated to accelerator research and development at DESY.

Here, we present a technique enabled by the PolariX TDS that allows for the tomographic reconstruction of the 5D $(x, x', y, y', t)$ phase space, where $x$, $x'$ and $y$, $y'$ are the transverse position and divergence and $t$ is time. The working principle consists of combining the streaking along various transverse directions with a transverse phase-space tomography based on a quadrupole scan \cite{jaster-merz:ipac22-mopopt021, jaster-merz:linac2022-mopori10}. The unique features of the 5D phase space tomography method, i.e., resolving complex phase-space distributions and their correlations, are studied through simulations based on the ARES beamline layout. The reconstruction accuracy is characterized for two different test distributions and the influence of the number of rotation and streaking angles on the reconstruction accuracy is investigated. 
The 5D phase-space information obtained with this method is useful to optimize and improve the beam quality of, e.g., free-electron lasers or beam driven plasma accelerators by detecting previously unnoticed correlations and other features in the distribution.
Furthermore, the reconstructed phase-space distribution is useful to benchmark existing simulation codes, or as an input to perform detailed simulation studies of advanced acceleration schemes.

\section{Working Principle}
The presented 5D tomography method consists of performing a 4D transverse phase-space tomography for each longitudinal slice of the bunch.
Here, the underlying principle of the tomography is to use lower-dimensional projections of an object along different angles to reconstruct its distribution in a higher-dimensional plane.
These higher-dimensional distributions are obtained by using a tomographic reconstruction algorithm such as the Filtered Back-projection \cite{kak2001principles}, the ART (Algebraic Reconstruction Technique) \cite{GORDON1970471}, the SART~(Simultaneous Algebraic Reconstruction Technique) \cite{ANDERSEN198481}, or the MENT (Maximum Entropy Tomography) \cite{MINERBO197948} algorithm. 

The transverse tomography of a particle bunch in an accelerator is performed by rotating the transverse phase spaces $(x,x')$ and $(y,y')$. This rotation is done by changing the phase advance $\mu_{x,y}$ experienced by the particle distribution.
It is especially useful to perform this in the normalized phase space, where for linear optics the phase advance $\mu_{x,y}$ is equivalent to the rotation angle $\theta_{x,y}$ of the phase spaces \cite{HOCK201136}. 
The conversion from real phase-space coordinates $x,y$ and $x',y'$ to normalized phase-space coordinates $x_{N},y_{N}$ and $x'_{N},y'_{N}$ is given by:
\begin{equation}
    \begin{pmatrix}
        x_{N} \\
        x'_{N} \\
        y_{N} \\
        y'_{N}
    \end{pmatrix}
    =
    \begin{pmatrix}
        \frac{1}{\sqrt{\beta_{x}}} & 0 & 0 & 0 \\
        \frac{\alpha_{x}}{\sqrt{\beta_{x}}} & \sqrt{\beta_{x}} & 0 & 0 \\
        0 & 0 & \frac{1}{\sqrt{\beta_{y}}} & 0 \\
        0 & 0 & \frac{\alpha_{y}}{\sqrt{\beta_{y}}} & \sqrt{\beta_{y}}
    \end{pmatrix}
    \begin{pmatrix}
        x \\
        x' \\
        y \\
        y'
    \end{pmatrix},
\end{equation}
where $\beta_{x,y}$ and $\alpha_{x,y}$ are the Courant-Snyder parameters \cite{COURANT19581}. To also resolve the correlations between the two transverse planes, their phase advances are controlled simultaneously.

The longitudinal information is obtained by streaking the bunch with the PolariX TDS along various transverse directions. This is required due to the overlap of the longitudinal information with the transverse one in the plane parallel to the streaking direction. Therefore, the full transverse distribution is only available when combining the projections of the bunch under various streaking directions. 

The working principle of the 5D phase-space tomography is depicted in Fig. \ref{fig:beamline}. 
For a fixed combination of transverse rotation angles $(\theta_{x},\,\theta_{y})$, the beam is streaked with the TDS, and its projection is recorded on a downstream screen. 
This is repeated for different streaking angles.
These screen images are used to reconstruct the 3D charge-density distribution $(x,y,t)$ of the bunch \cite{Marx_2017, Marx2019, Marx:427780, Marchetti2021}. 
The transverse profiles are reconstructed at the location of the screen. 
The longitudinal information is reconstructed at the TDS center. 
This assumes that the longitudinal distribution does not change between the two locations which is valid as long as the longitudinal displacement of a particle does not exceed the longitudinal resolution of the reconstruction. 
The reconstruction procedure is repeated for all desired phase advance combinations.
Each 3D reconstruction can be regarded as the projection of the $(x,x',y,y',t)$ phase space onto the $(x,y,t)$ plane for different transverse phase advance combinations.
To reconstruct the higher-dimensional distribution, the 3D reconstructions are combined. This is done for each longitudinal slice $t_{s}$ individually.
Using the reconstructed sliced projections $(x,y)_{t_{s}}$, the transverse charge-density distribution $(x,x',y,y')$ can be reconstructed for each slice $t_{s}$, similar to the reconstruction of the 4D transverse charge density in \cite{HOCK20138, PhysRevAccelBeams.23.032804}. This distribution is reconstructed at a location upstream of the TDS and of all the quadrupoles that are used to scan the phase advance.
By combining all longitudinal slices, the 5D charge-density distribution is obtained.

To obtain the optimal beamline settings for the 5D tomography, several factors have to be considered. 
First, the setup should allow for accurate tuning of the transverse phase advance in both planes in a range of \SI[]{180}{\degree} to obtain a good sampling of the transverse distribution. 
Second, the longitudinal resolution of the reconstruction should be sufficient to identify all features of interest. This longitudinal resolution $R$ is given by \cite{Marx_2017_TDS}
\begin{equation}
    \label{TDS_resolution}
    R= \frac{c p \sigma^{\text{off}}} {2 \pi e f V L}, 
\end{equation}
where $\sigma^{\text{off}}$ is the transverse spot size at the screen when the TDS is switched off, $p$ is the average momentum of the bunch, $c$ the speed of light in vacuum, $e$ the elementary charge, $f$ the TDS X-band working frequency, $V$ the peak voltage, and $L$ the drift length between the TDS center and the downstream screen. 
If transverse-longitudinal correlations are present in the distribution, the projected transverse spot size can be larger than the spot size of each individual slice. 
The longitudinal resolution could therefore be improved by determining the resolution using the transverse spot size of each slice. 
These spot sizes of the individual slices can be obtained from the screen image of the streaked bunch. 
Since the spot size along the streaking direction is the one of interest, the screen image streaked perpendicular to the streaking angle of interest has to be considered.
With this approach, also a variable longitudinal resolution along the bunch would be possible. 
The transverse spot size is directly linked to the beta functions at the screen. These need to be matched to a constant value per plane in order to obtain a uniform longitudinal resolution for all transverse phase advance combinations. For a fixed TDS setting, the final longitudinal resolution is given by the largest transverse spot size of all streaking directions for all phase advance combinations. 
Furthermore, the screen resolution needs to be taken into account to ensure that the pixel size can resolve the features of the projected bunch.

The longitudinal information of the distribution is obtained by converting the screen coordinate in the direction of the streaking angle $u^{*}$ to the longitudinal position along the bunch $\Delta t$ according to
\begin{equation}
    \Delta t = \frac{u^{*}}{c S},
\end{equation}
where $S$ is the shear parameter defined as \cite{PhysRevSTAB.12.050704}
\begin{equation}
    S = \frac{2 \pi e f V L}{c^{2}p}.
\end{equation}  

When measuring the longitudinal bunch information with a TDS, typically, screen images are recorded at both RF zero crossings to account for transverse-longitudinal correlations within the bunch. Here, screen images are recorded only at the first zero crossing to keep a feasible measurement and analysis time.

\begin{table*}
	\centering
	\caption{Beam Parameters of the Original and Reconstructed Distributions}
    \begin{ruledtabular}
		\begin{tabular}{ll ll ll}
				& & \multicolumn{2}{c}{\textbf{Gaussian with correlations}} & \multicolumn{2}{c}{\textbf{Multi beam structure}}\\
			\textbf{Parameter} & \textbf{Units}  & {Original} & {Reconstruction} & {Original} & {Reconstruction}\\
			\colrule
			$E$ & \si[]{MeV} & 155 & - & 155 & - \\
			$\sigma_{E}$ & \si[]{\percent} & 0.1 & -  & 0.1 & -\\
			$Q$ & \si[]{pC} & 1 & - & 1 & -\\
			$\sigma_{\tau}$ & \si[]{fs} & 200.00 & 200.74 & 200.01 & 205.43\\
			$\epsilon_{x}$ & \si[]{\metre \radian} & \SI[]{3.30 e-9}{} & \SI[]{3.38 e-9}{} & \SI[]{15.67 e-9}{} & \SI[]{18.75 e-9}{}\\
			$\epsilon_{y}$ & \si[]{\metre \radian} & \SI[]{3.31 e-9}{} & \SI[]{3.48 e-9}{}& \SI[]{14.70 e-9}{} & \SI[]{18.25 e-9}{}\\
			$\epsilon_{x}^{n}$ & \si[]{\micro\metre} & \SI[]{1.00}{} & \SI[]{1.02}{}& \SI[]{4.75}{} & \SI[]{5.69}{}\\
			$\epsilon_{y}^{n}$ & \si[]{\micro\metre} & \SI[]{1.01}{} & \SI[]{1.06}{}& \SI[]{4.46}{} & \SI[]{5.54}{}\\
			$\alpha_{x}$ & - & 0.00 & 0.00 & 0.00 & 0.00\\
			$\alpha_{y}$ & - & 0.00 & 0.00 & 0.00 & 0.00\\
			$\beta_{x}$ & \si[]{\metre} & 5.00 & 4.97 & 5.01 & 5.14\\
			$\beta_{y}$ & \si[]{\metre} & 5.00 & 4.96 & 5.01 & 5.23\\
		\end{tabular}
	\label{tab:beam_parameters}
    \end{ruledtabular}
\end{table*}
\begin{figure*}
    \centering
    \includegraphics*[width=1.\textwidth]{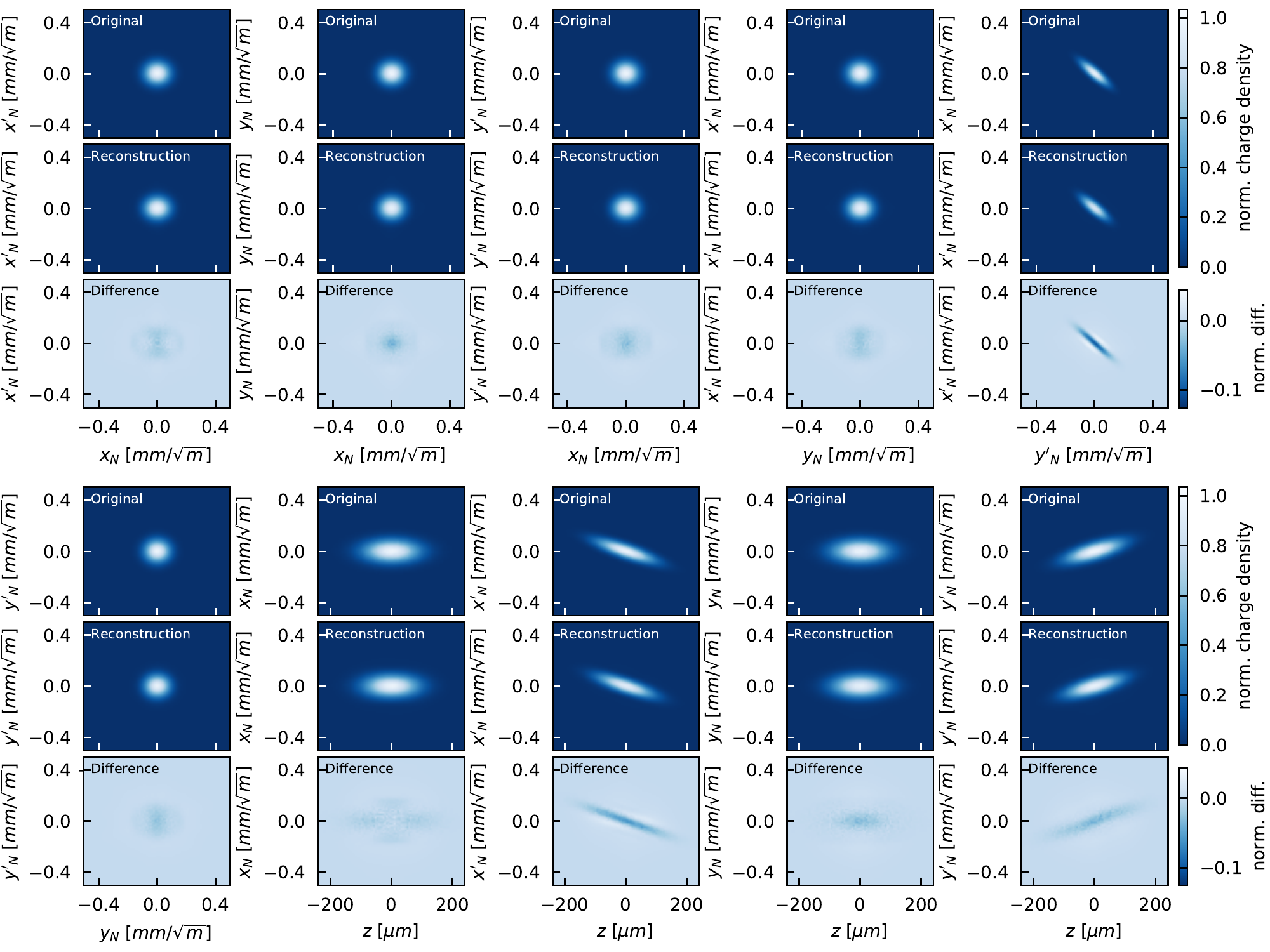}
    \caption{All ten 2D projections of the 5D phase-space distribution of the Gaussian distribution with correlations. The top rows show the projection of the original distribution, the middle rows show the projections of the reconstructed phase-space distribution, and the bottom rows show the difference between the two. All projections are displayed in normalized transverse phase space and the density is normalized to the maximum value of the corresponding 2D original projection.}
    \label{fig:all_2D_reconstructions}
\end{figure*}

\begin{figure*}
    \centering
    \includegraphics*[width=1.\textwidth]{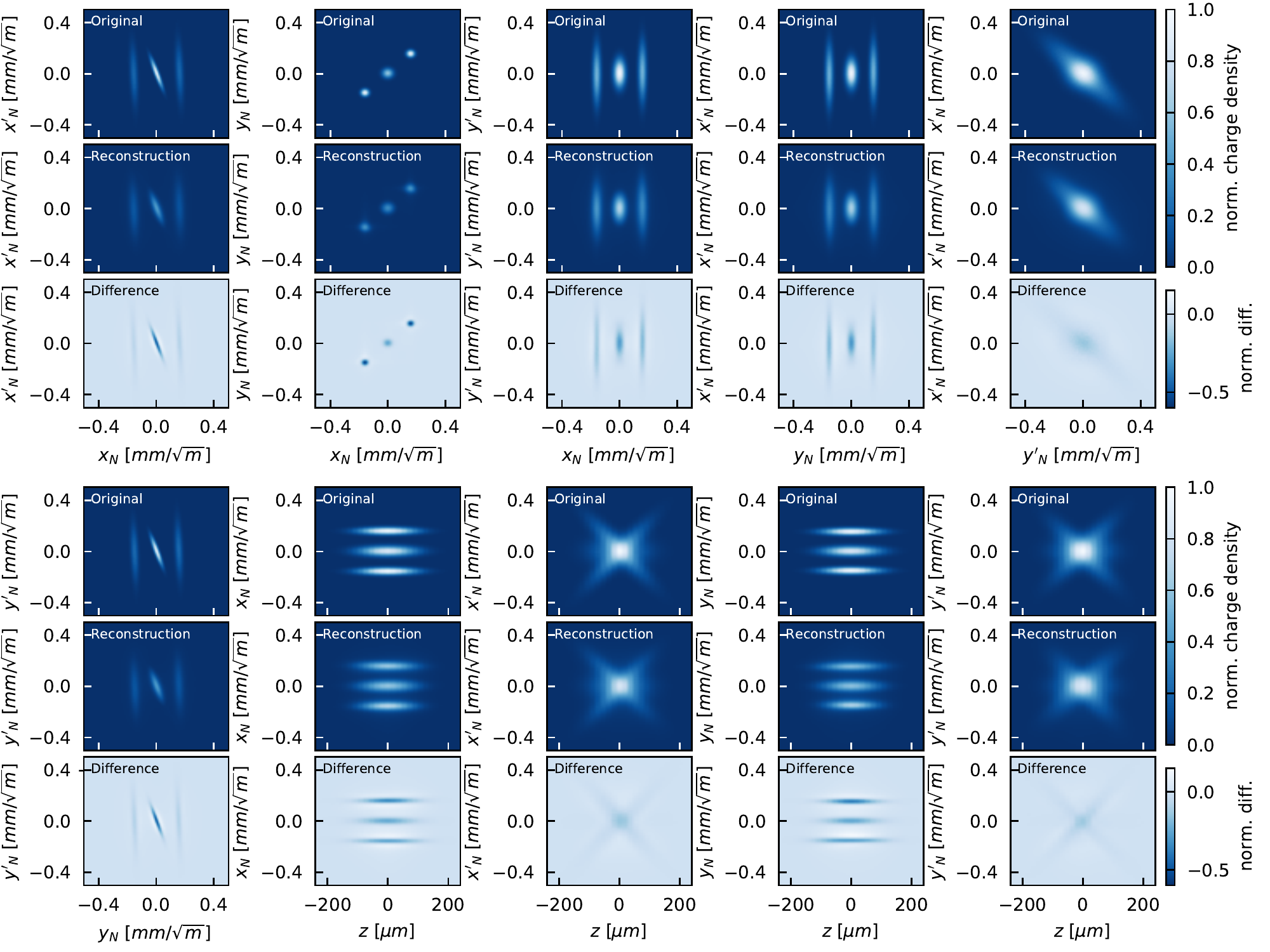}
    \caption{All ten 2D projections of the 5D phase-space distribution of the multi-beam distribution. The top rows show the projection of the original distribution, the middle rows show the projections of the reconstructed phase-space distribution, and the bottom rows show the difference between the two. All projections are displayed in normalized transverse phase space and the density is normalized to the maximum value of the corresponding 2D original projection.}
    \label{fig:all_2D_reconstructions_multi_beam}
\end{figure*}
\section{Proof-of-principle simulation studies}
\label{demonstration_of_method}

The present study investigates the capabilities of the 5D tomography method to reconstruct phase-space densities with correlations and complex features.
This is done by studying two different distributions in simulation studies and comparing the known input distribution to the reconstruction result. 
For these studies, the ARES beamline layout shown in Fig. \ref{fig:beamline} is used, which consists of five quadrupoles and two PolariX TDSs.
The first distribution is based on realistic beam parameters at ARES. This distribution is a Gaussian distribution with imprinted correlations in the $(x'\text{--}z)$, $(y'\text{--}z)$ and $(x'\text{--}y')$ planes. It is depicted in Fig. \ref{fig:all_2D_reconstructions} with the label ``original''. 
The second distribution is used to test the ability of the method to reconstruct arbitrary phase-space distributions.
It consists of three superimposed Gaussian beams with transverse offsets with respect to each other and longitudinal correlations in the transverse momenta, resulting in a complex phase-space structure.
This distribution is depicted in Fig. \ref{fig:all_2D_reconstructions_multi_beam} with the label ``original''. 
Both distributions are generated in \textsc{ocelot} \cite{AGAPOV2014151, tomin:ipac17-wepab031} and feature the same initial Courant-Snyder parameters, which allows for the use of the same quadrupole settings for the two distributions. 
The second distribution features a larger transverse emittance due to its multi-bunch structure. All the initial beam parameters are listed in Table~\ref{tab:beam_parameters}, where $E$ is the average energy, $\sigma_{E}$ the RMS energy spread, $Q$ the charge, $\sigma_{t}$ the RMS bunch duration, and $\epsilon_{x,y}$ ($\epsilon_{x,y}^{n}$) the geometric (normalized) RMS emittance.

The five quadrupoles and one of the PolariX TDSs are used to achieve the desired phase advances and streaking angles. 
To determine the longitudinal resolution of the reconstructed distribution, the maximum unstreaked spot size at the screen is considered. It is obtained by tracking the beam distribution through the beamline for all phase-advance combinations with the TDS switched off. The maximum RMS transverse spot size is then determined by comparing all transverse projections onto all streaking planes.
From Eq. \ref{TDS_resolution}, a peak voltage $V$ of \SI{3.6}{MV} is required to achieve a longitudinal resolution of \SI{20}{fs}, with an X-band TDS frequency of \SI{11.992}{GHz}, a maximum spot size $\sigma^{\text{off}}$ of \SI{247}{\micro\metre}, and a drift $L$ of \SI{7.21}{m}.
In practice, each of the available PolariX TDSs would be able to streak the beam with up to \SI{20}{MV} peak voltage. By using both TDSs with \SI{20}{MV} each, assuming the same unstreaked spot size and the drift length of $L = \SI{6.67}{m}$, a resolution of \SI{1.91}{fs} could be achieved. 

In this study, only the first TDS in Fig. \ref{fig:beamline} is used to achieve the desired \SI{20}{fs}, and a constant shear parameter for all phase-advance settings and streaking angles is assumed. 
The streaking angle of this TDS is varied over a range of \SI[]{180}{\degree} in 40 steps. The quadrupole strengths are set to scan the transverse phase advances over a range of \SI{180}{\degree} in 40 steps. The projections of the distributions are recorded on the screen station downstream of the TDS. The simulated screen has a size of \SI[]{2.02 x 2.02}{\cm} to fit \SI[]{\pm 4}{\sigma} of the streaked RMS bunch duration, and \SI[product-units = single]{2001 x 2001}{pixels}, which corresponds to a resolution in the range of the ARES screen stations. Moreover, it ensures a minimum of \SI{4}{pixel} per transverse RMS spot size and more than \SI{8}{pixel} per transverse spot size for \SI{97}{\percent} of the scanned phase-advance combinations. For the reconstruction a region of interest of \SI{201}{pixels} is selected in the transverse direction which fits \SI[]{\pm 4}{\sigma} of the transverse RMS beam size. The reconstruction of the transverse distribution is performed at the location of the screen station upstream of the first used quadrupole.

All simulations are performed in \textsc{ocelot} \cite{AGAPOV2014151, tomin:ipac17-wepab031} using distributions with \SI[]{4000000}{particles}. Second-order transfer maps are used for all elements except the TDS, where only a first-order transfer map is available. 
Space-charge effects are not taken into account because according to the laminarity parameter defined in \cite{Ferrario2014, PhysRevSTAB.5.014201}, the bunch is emittance dominated throughout the beamline.
This is confirmed by simulations including space-charge forces for the phase-advance setting with the maximal laminarity parameter along the beamline. The tracked distribution is compared to the corresponding simulation without space charge for vertical and horizontal streaking. No significant differences between the two cases were observed and therefore space-charge effects are neglected in the presented studies.

The final transverse reconstruction is performed in normalized phase space. For this, the reconstructed sliced $(x, y)$ projections at the screen downstream of the PolariX TDS are normalized using the beta functions obtained by propagating the initial Courant-Snyder parameters through the beamline for each phase-advance combination. The final reconstruction, with the transverse distribution obtained at the screen station upstream of the first used quadrupole and the longitudinal information obtained at the TDS center, has a size of \SI{201}{bins} in all transverse dimensions and \SI{80}{bins} in the longitudinal plane, which, respectively, results in a resolution of \SI{5}{ \micro \meter / \sqrt{m}} and \SI{20}{fs}. 
All tomographic reconstructions in this study use a Python scikit-image \cite{van2014scikit} implementation of the SART algorithm \cite{ANDERSEN198481}. 
This algorithm uses an iterative solver to obtain the reconstruction. A good reconstruction is already obtained in a single iteration. Increasing the number of iterations improves the reconstruction of sharp features in the distribution at the cost of an increased noise level \cite{ANDERSEN198481}.
For each tomographic reconstruction in this study, two iterations over the algorithm are performed, where the second iteration uses the first reconstruction as an input for the initial reconstruction estimate. A condition is imposed in the algorithm to ensure charge invariance and strictly positive charge-density values.

\begin{figure}
    \centering
    \includegraphics*[width=1.\columnwidth]{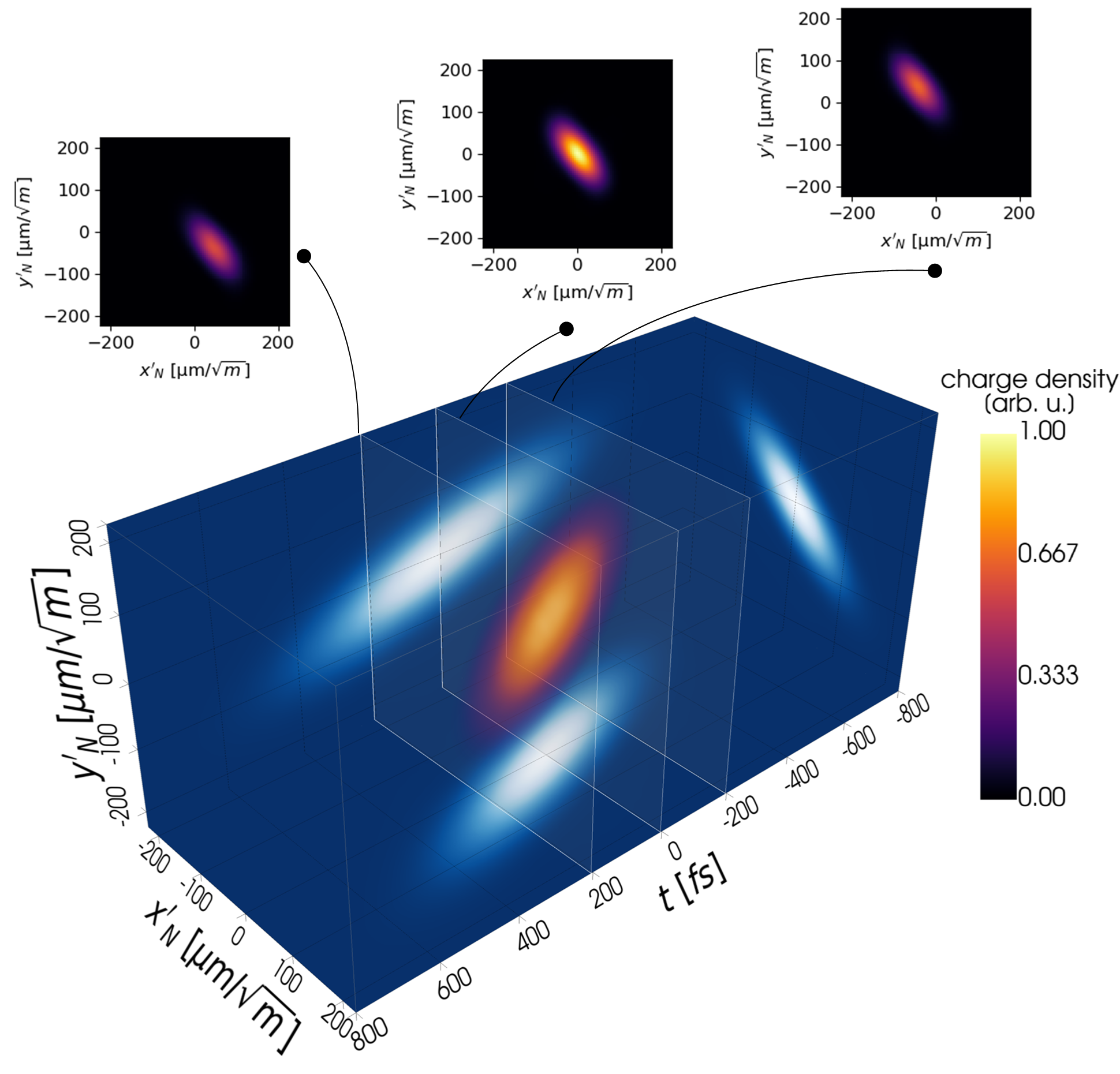}
    \caption{3D render of the reconstructed 5D phase space of the Gaussian distribution with correlations. The distribution is projected onto the normalized $(x'_{N}, y'_{N}, t)$ phase space. The charge density is normalized to the maximum value of the 3D distribution. The transverse $(x'_{N}, y'_{N})$ central slice and the slices $\pm$ \SI{200}{fs} from the center are shown exemplarily. In addition, the 2D projections of the 3D distribution are shown in a blue color coding. The 3D render is done with pyvista \cite{sullivan2019pyvista}, a Python implementation of VTK \cite{schroeder2006visualization}.}
    \label{fig:render_xp_yp}
\end{figure}

\begin{figure}
    \centering
    \includegraphics*[width=1.\columnwidth]{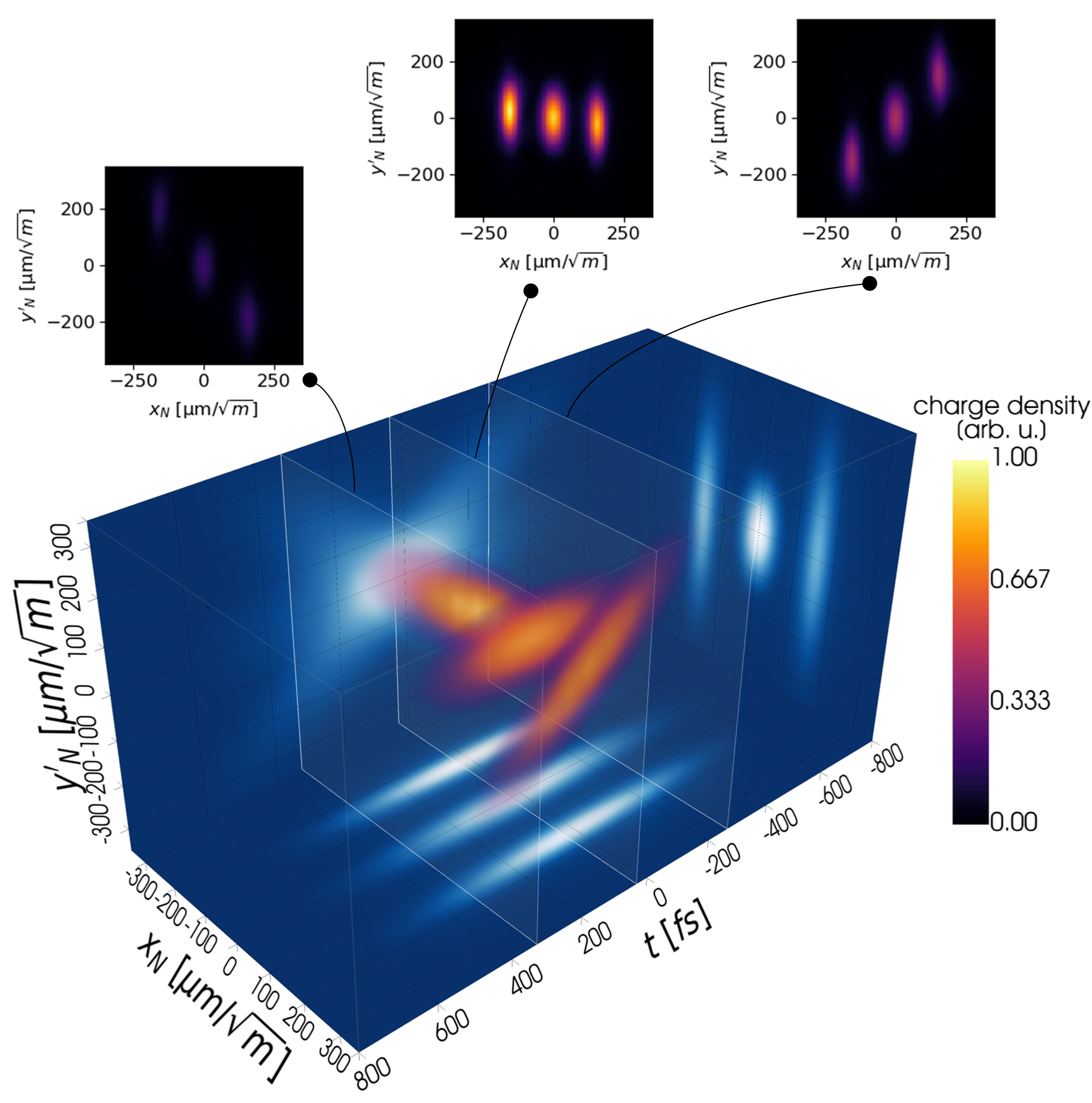}
    \caption{3D render of the reconstructed 5D phase space of the multi-beam distribution. The distribution is projected onto the $(x_{N}, y'_{N}, t)$ phase space. The charge density is normalized to the maximum value of the 3D distribution. The transverse $(x_{N}, y'_{N})$ slices at \SI{-246}{fs}, \SI{62}{fs} and \SI{369}{fs} are shown exemplarily. In addition, the 2D projections of the 3D distribution are shown in a blue color coding. The 3D render is done with pyvista \cite{sullivan2019pyvista}, a Python implementation of VTK \cite{schroeder2006visualization}.}
    \label{fig:render_xp_yp_sub_structure}
\end{figure}

\section{Reconstruction results}
\label{results}
For the first distribution of a Gaussian beam with imprinted correlations, the reconstructed 5D phase space is shown in Fig. \ref{fig:render_xp_yp} as a projection onto the $(x'_{N}, y'_{N}, t)$ phase space.
The reconstructed Courant-Snyder parameters, emittances, and bunch lengths are listed in Table~\ref{tab:beam_parameters}. 
Tomographic methods such as the SART tend to introduce spurious charge density even far off-axis where no charge should be present \cite{Marchetti2021} especially when only a limited number ($<$ 100) of rotation angles are used. 
Therefore, to analyze the emittances and Courant-Snyder parameters of the reconstructed distributions only values within a \SI[]{\pm 3}{\sigma} range from the bunch center are considered. The bunch duration $\sigma_{\tau}$ is calculated as the standard deviation of the distribution projected onto the time axis.
Excellent agreement with relative discrepancies of $\lesssim$ \SI{5}{\percent} is achieved for the beam parameters. 
The imprinted correlations in the distribution are analyzed in normalized phase space and quantified by the slope of a linear fit which is applied to the corresponding 2D projections of the phase space. The original and reconstructed slopes are listed in Table \ref{tab:correlations} and show relative discrepancies below \SI{5}{\percent}.

\begin{table}
	\centering
	\caption{Original and Reconstructed Correlations of the Gaussian Distribution with Correlations}
    \begin{ruledtabular}
		\begin{tabular}{lll}
            \textbf{Plane} & \textbf{Original}  & \textbf{Reconstruction} \\
			\colrule
            $(x'_{N}\text{--}y'_{N})$ & \SI{-0.848}{} & \SI{-0.814}{} \\
            $(x'_{N}\text{--}z)$ & \SI{-0.835}{m^{1/2}} & \SI{-0.793}{m^{1/2}} \\
            $(y'_{N}\text{--}z)$ & \SI{0.706}{m^{1/2}} & \SI{0.677}{m^{1/2}} \\
        \end{tabular}
        \label{tab:correlations}
    \end{ruledtabular}
\end{table}

Additionally, all ten 2D projections of the 5D distribution are shown in Fig. \ref{fig:all_2D_reconstructions} and compared to the original input distribution. The projections are normalized to the maximum value of the projection of the original distribution and the normalized difference between original and reconstructed projection are shown. All ten projections are reconstructed with excellent agreement with a maximum relative discrepancy of \SI{13}{\percent}. The largest discrepancy appears in the $(x'_{N}, y'_{N})$ projection where generally a widening of the reconstructed distribution is observed.

The second distribution has a complex phase-space distribution and therefore a larger maximal unstreaked transverse RMS spot size at the screen of \SI{722.07}{\micro\metre}. As a result, some adjustments are required compared to the settings described in Section \ref{demonstration_of_method}. Using the same TDS and quadrupole settings as for the first distribution, a longitudinal resolution of \SI{57.21}{fs} is achieved, and 27 longitudinal slices are reconstructed. 
Due to the bigger spot size, a larger region of interest of \SI{431}{bins} is selected for the reconstruction, which fits \SI[]{\pm 3}{\sigma} of the maximum RMS spot size. 
All other parameters remain unchanged with respect to the reconstruction of the first distribution.

The reconstructed bunch duration and Courant-Snyder parameters for the multi-structure beam agree well with the original parameters with discrepancies of around \SI{4}{\percent}. 
However, the reconstructed emittances show large deviations of up to \SI{24}{\percent}. 
Possible reasons could be an insufficient number of projection angles and screen resolution to resolve the sharp and thin features of the distribution. 
Due to the substantial increase in required computational power and memory when simulating many projection angles and a high screen resolution, this is not investigated further.
The reconstructed 5D phase space is shown in Fig. \ref{fig:render_xp_yp_sub_structure} as a projection onto the $(x_{N}, y'_{N}, t)$ phase space which exhibits a more interesting structure than the previously shown $(x'_{N}, y'_{N}, t)$ phase space. The substructure consisting of three sub-beams is well reconstructed and clearly visible. All 2D projections of the 5D phase-space distribution are shown in Fig \ref{fig:all_2D_reconstructions_multi_beam}.
Also here the complex substructure of the bunch is recovered and the reconstruction of the detailed features can be appreciated. 
The projections onto the $(x'_{N},y'_{N})$, $(x'_{N},z)$, and $(y'_{N}, z)$ phase spaces are accurately reconstructed and show relative discrepancies below \SI{20}{\percent}.
The largest relative discrepancies of up to \SI{60}{\percent} appear in the $(x_{N},x'_{N})$, $(y_{N},y'_{N})$, and $(x_{N},y_{N})$ projections where a blurring of the reconstructed phase space is observed.
However, this blurring does not prevent a precise qualitative description of the phase space.
Therefore, this result demonstrates the method's potential to reconstruct the 5D phase-space distribution of complex distributions.
\begin{figure}
    \centering
    \includegraphics*[width=1.\columnwidth]{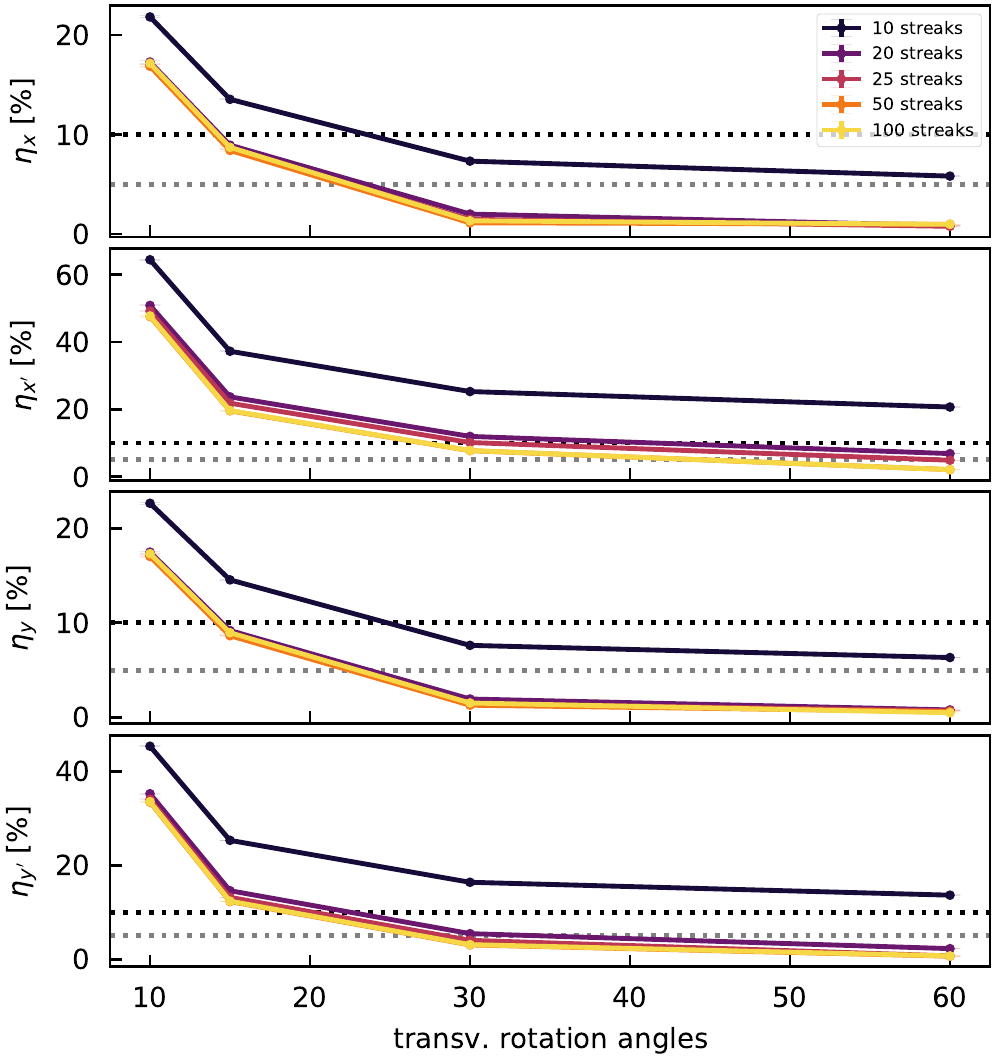}
    \caption{Relative discrepancies, as defined in Eq. \ref{reconstruction_discrepancy}, between the reconstructed and input distribution of all transverse planes for various numbers of transverse rotation and streaking angles. The dotted gray and black lines show the \SI{5}{\percent} and \SI{10}{\percent} levels, respectively.}
    \label{fig:angle_scan}
\end{figure}

\section{Influence of rotation angles}
The influence of the number of transverse rotation angles, as well as streaking angles, on the reconstruction accuracy is studied in this section.
This is done to find an optimal balance between the required preparatory simulation and later experimental measurement time, which is determined by the number of scanned angles, and the final accuracy of the reconstruction.
For this, several combinations of rotation angles and streaking angles are compared.
The same Gaussian distribution with correlations as studied in Section \ref{demonstration_of_method} and \ref{results} is used.
The tracking is performed for 60 transverse rotation angles for both transverse planes and 100 streaking angles. The tomographic reconstruction is performed using different sub-datasets to study a variety of rotation and streaking angle combinations without having to repeat the computational expensive tracking which runs approximately \SI{42}{\hour} on 6 AMD EPYC 75F3 nodes on a computing cluster. As a measure for the accuracy of the reconstruction, the relative discrepancies $\eta_{i,j} = (\sigma_{i,j}^{recon} - \sigma_{i,j}^{input}) / \sigma_{i,j}^{input}$ are computed for each longitudinal slice $j$ and transverse 1D projection of the reconstruction, where $i=x,x',y,y'$. The values $\sigma_{i,j}^{recon}$ are obtained from the standard deviation of a Gaussian fit to the reconstructed distribution, the values $\sigma_{i,j}^{input}$ are obtained from the standard deviation of the input distribution to be less sensitive to a low number of particles in some slices. The value $\eta_{i,j}$ of each slice is weighted by the corresponding charge $Q_{j}$ of the slice, and the $\eta_{i,j}$ values are summed for each transverse plane $i$ individually:
\begin{equation}
    \label{reconstruction_discrepancy}
    \eta_{i} = \sum_{j} \eta_{i,j} Q_{j} .
\end{equation}

Figure \ref{fig:angle_scan} shows the discrepancies for the four transverse planes and various combinations of rotation and streaking angles. The reconstruction accuracy improves with increasing number of rotation and streaking angles. Increasing the number of streaking angles from 50 to 100 results in no significant improvement of the reconstruction accuracy which in this case is dominated by the number of transverse rotation angles.
To obtain discrepancies of $\lesssim$ \SI{5}{\percent} in all reconstructed planes, $\sim$ 50 transverse rotation angles and $\sim$ 50 streaking angles are required. To obtain discrepancies of $\lesssim$ \SI{10}{\percent} in all planes, $\sim$ 30 transverse rotation angles and $\sim$ 25 streaking angles are required. 
In both cases this is limited by the accuracy of the reconstruction in the $x'$ plane. 
This is probably due to the $(x'- t)$ correlation in the distribution and the resulting smaller RMS slice divergences. 
Furthermore, a difference between the $x'$ and $y'$ planes is observed.
Again, this is likely caused by the correlations in these planes which are larger in $(x'- t)$ than in $(y'-t)$.
For a pure Gaussian distribution without correlations, no significant differences between the accuracy of the $x$, $x'$ planes and $y$, $y'$ planes is observed. 

The relative discrepancy of the bunch duration is obtained by comparing the standard deviation of the input distribution to the standard deviation of a Gaussian fit of the 1D projection of the reconstructed distribution onto $t$. It is unaffected by the number of transverse rotation and streaking angles and the average relative discrepancy is \SI[separate-uncertainty = true,multi-part-units=single]{0.273(4)}{\percent}. The independence on the number of rotation angles can be explained by the imposed constraint of the method to preserve the charge of each longitudinal slice once the 3D distribution $(x, y, t)$ is reconstructed. 
No impact of the number of streaking angles on the accuracy of the bunch duration reconstruction is observed.

The presented study uses the SART algorithm for the tomographic reconstruction. The 5D tomography method itself, however, is independent of the used reconstruction algorithm. 
Therefore, one possible approach to further reduce the required transverse rotation and streaking angles is to explore other reconstruction algorithms, such as the MENT algorithm \cite{MINERBO197948, scheins2004tomographic}.
Recent studies \cite{PhysRevAccelBeams.25.122803, PhysRevLett.130.145001} using a machine learning approach to reconstruct the phase space show promising results and their application to the 5D tomography method could be explored. 

\section{CONCLUSION}
The presented tomographic method enables the accurate reconstruction and visualization of the full 5D $(x, x', y, y', t)$ phase-space distribution of accelerated bunches. Simulation studies for a Gaussian distribution with a longitudinally correlated transverse momentum offset show excellent agreement with the original beam parameters and correlations with discrepancies $\lesssim$ \SI{5}{\percent}.
For a complex, multi-beam phase-space distribution, all features are clearly reconstructed but are more smeared out. This is expected to stem from the thin and sharp features of the distribution which require a larger number of rotation angles and a small pixel size of the recording screen to be reproduced accurately.
In general, an improved reconstruction accuracy is achieved by increasing the number of transverse rotation and streaking angles. To improve the accuracy of the method for small numbers of rotation and streaking angles, different reconstruction algorithms could be explored. 
Experimental measurements are planned in the near future to demonstrate the practical applicability of the method.

\begin{acknowledgments}
We would like to thank B. Marchetti for the suggestion of studying the 5D phase-space tomography with the PolariX TDS and the PolariX collaboration for helpful discussions on the device. We would also like to thank A. Ferran Pousa for many helpful discussions on beam dynamics and the efficient computational implementation of the method, D. Marx and P. Gonzalez Caminal for helpful discussions on beam dynamics in the PolariX TDS, and W. Kuropka for providing the implementation of the ARES beamline in \textsc{ocelot}. Finally, we thank A. Wolski for helpful discussions on the 4D tomography method and inspiring discussions on the 5D simulation studies. This research was supported in part through the Maxwell computational resources operated at Deutsches Elektronen-Synchrotron DESY, Hamburg, Germany. We acknowledge support from DESY (Hamburg, Germany), a member of the Helmholtz Association HGF.
\end{acknowledgments}

\bibliography{5D_tomography}

\begin{thebibliography}{45}%
\makeatletter
\providecommand \@ifxundefined [1]{%
 \@ifx{#1\undefined}
}%
\providecommand \@ifnum [1]{%
 \ifnum #1\expandafter \@firstoftwo
 \else \expandafter \@secondoftwo
 \fi
}%
\providecommand \@ifx [1]{%
 \ifx #1\expandafter \@firstoftwo
 \else \expandafter \@secondoftwo
 \fi
}%
\providecommand \natexlab [1]{#1}%
\providecommand \enquote  [1]{``#1''}%
\providecommand \bibnamefont  [1]{#1}%
\providecommand \bibfnamefont [1]{#1}%
\providecommand \citenamefont [1]{#1}%
\providecommand \href@noop [0]{\@secondoftwo}%
\providecommand \href [0]{\begingroup \@sanitize@url \@href}%
\providecommand \@href[1]{\@@startlink{#1}\@@href}%
\providecommand \@@href[1]{\endgroup#1\@@endlink}%
\providecommand \@sanitize@url [0]{\catcode `\\12\catcode `\$12\catcode
  `\&12\catcode `\#12\catcode `\^12\catcode `\_12\catcode `\%12\relax}%
\providecommand \@@startlink[1]{}%
\providecommand \@@endlink[0]{}%
\providecommand \url  [0]{\begingroup\@sanitize@url \@url }%
\providecommand \@url [1]{\endgroup\@href {#1}{\urlprefix }}%
\providecommand \urlprefix  [0]{URL }%
\providecommand \Eprint [0]{\href }%
\providecommand \doibase [0]{https://doi.org/}%
\providecommand \selectlanguage [0]{\@gobble}%
\providecommand \bibinfo  [0]{\@secondoftwo}%
\providecommand \bibfield  [0]{\@secondoftwo}%
\providecommand \translation [1]{[#1]}%
\providecommand \BibitemOpen [0]{}%
\providecommand \bibitemStop [0]{}%
\providecommand \bibitemNoStop [0]{.\EOS\space}%
\providecommand \EOS [0]{\spacefactor3000\relax}%
\providecommand \BibitemShut  [1]{\csname bibitem#1\endcsname}%
\let\auto@bib@innerbib\@empty
\bibitem [{\citenamefont {Ferrario}\ \emph {et~al.}(2014)\citenamefont
  {Ferrario}, \citenamefont {Migliorati},\ and\ \citenamefont
  {Palumbo}}]{Ferrario2014}%
  \BibitemOpen
  \bibfield  {author} {\bibinfo {author} {\bibfnamefont {M.}~\bibnamefont
  {Ferrario}}, \bibinfo {author} {\bibfnamefont {M.}~\bibnamefont
  {Migliorati}},\ and\ \bibinfo {author} {\bibfnamefont {L.}~\bibnamefont
  {Palumbo}},\ }\href {https://doi.org/10.5170/CERN-2014-009.331} {\bibinfo
  {title} {Space charge effects}} (\bibinfo {year} {2014})\BibitemShut
  {NoStop}%
\bibitem [{\citenamefont {Dohlus}\ \emph {et~al.}(2006)\citenamefont {Dohlus},
  \citenamefont {Limberg},\ and\ \citenamefont {Emma}}]{dohlus2006bunch}%
  \BibitemOpen
  \bibfield  {author} {\bibinfo {author} {\bibfnamefont {M.}~\bibnamefont
  {Dohlus}}, \bibinfo {author} {\bibfnamefont {T.}~\bibnamefont {Limberg}},\
  and\ \bibinfo {author} {\bibfnamefont {P.}~\bibnamefont {Emma}},\ }\bibfield
  {title} {\bibinfo {title} {{Bunch Compression for Linac-Based {FEL}'s.
  Electron Bunch Length Compression}},\ }\href
  {https://www.osti.gov/biblio/877227} {\bibfield  {journal} {\bibinfo
  {journal} {ICFA Beam Dynamics Newsletter}\ }\textbf {\bibinfo {volume} {38}}
  (\bibinfo {year} {2006})}\BibitemShut {NoStop}%
\bibitem [{\citenamefont {Zheng}\ \emph {et~al.}(2019)\citenamefont {Zheng},
  \citenamefont {Shao}, \citenamefont {Du}, \citenamefont {Power},
  \citenamefont {Wisniewski}, \citenamefont {Liu}, \citenamefont {Whiteford},
  \citenamefont {Conde}, \citenamefont {Doran}, \citenamefont {Jing},
  \citenamefont {Tang},\ and\ \citenamefont
  {Gai}}]{PhysRevAccelBeams.22.072805}%
  \BibitemOpen
  \bibfield  {author} {\bibinfo {author} {\bibfnamefont {L.}~\bibnamefont
  {Zheng}}, \bibinfo {author} {\bibfnamefont {J.}~\bibnamefont {Shao}},
  \bibinfo {author} {\bibfnamefont {Y.}~\bibnamefont {Du}}, \bibinfo {author}
  {\bibfnamefont {J.~G.}\ \bibnamefont {Power}}, \bibinfo {author}
  {\bibfnamefont {E.~E.}\ \bibnamefont {Wisniewski}}, \bibinfo {author}
  {\bibfnamefont {W.}~\bibnamefont {Liu}}, \bibinfo {author} {\bibfnamefont
  {C.~E.}\ \bibnamefont {Whiteford}}, \bibinfo {author} {\bibfnamefont
  {M.}~\bibnamefont {Conde}}, \bibinfo {author} {\bibfnamefont
  {S.}~\bibnamefont {Doran}}, \bibinfo {author} {\bibfnamefont
  {C.}~\bibnamefont {Jing}}, \bibinfo {author} {\bibfnamefont {C.}~\bibnamefont
  {Tang}},\ and\ \bibinfo {author} {\bibfnamefont {W.}~\bibnamefont {Gai}},\
  }\bibfield  {title} {\bibinfo {title} {Experimental demonstration of the
  correction of coupled-transverse-dynamics aberration in an rf
  photoinjector},\ }\href {https://doi.org/10.1103/PhysRevAccelBeams.22.072805}
  {\bibfield  {journal} {\bibinfo  {journal} {Phys. Rev. Accel. Beams}\
  }\textbf {\bibinfo {volume} {22}},\ \bibinfo {pages} {072805} (\bibinfo
  {year} {2019})}\BibitemShut {NoStop}%
\bibitem [{\citenamefont {Dowell}\ \emph {et~al.}(2018)\citenamefont {Dowell},
  \citenamefont {Zhou},\ and\ \citenamefont
  {Schmerge}}]{PhysRevAccelBeams.21.010101}%
  \BibitemOpen
  \bibfield  {author} {\bibinfo {author} {\bibfnamefont {D.~H.}\ \bibnamefont
  {Dowell}}, \bibinfo {author} {\bibfnamefont {F.}~\bibnamefont {Zhou}},\ and\
  \bibinfo {author} {\bibfnamefont {J.}~\bibnamefont {Schmerge}},\ }\bibfield
  {title} {\bibinfo {title} {Exact cancellation of emittance growth due to
  coupled transverse dynamics in solenoids and rf couplers},\ }\href
  {https://doi.org/10.1103/PhysRevAccelBeams.21.010101} {\bibfield  {journal}
  {\bibinfo  {journal} {Phys. Rev. Accel. Beams}\ }\textbf {\bibinfo {volume}
  {21}},\ \bibinfo {pages} {010101} (\bibinfo {year} {2018})}\BibitemShut
  {NoStop}%
\bibitem [{\citenamefont {de~Loos}\ \emph {et~al.}(2006)\citenamefont
  {de~Loos}, \citenamefont {van~der Geer}, \citenamefont {Saveliev},
  \citenamefont {Pavlov}, \citenamefont {Reitsma}, \citenamefont {Wiggins},
  \citenamefont {Rodier}, \citenamefont {Garvey},\ and\ \citenamefont
  {Jaroszynski}}]{PhysRevSTAB.9.084201}%
  \BibitemOpen
  \bibfield  {author} {\bibinfo {author} {\bibfnamefont {M.~J.}\ \bibnamefont
  {de~Loos}}, \bibinfo {author} {\bibfnamefont {S.~B.}\ \bibnamefont {van~der
  Geer}}, \bibinfo {author} {\bibfnamefont {Y.~M.}\ \bibnamefont {Saveliev}},
  \bibinfo {author} {\bibfnamefont {V.~M.}\ \bibnamefont {Pavlov}}, \bibinfo
  {author} {\bibfnamefont {A.~J.~W.}\ \bibnamefont {Reitsma}}, \bibinfo
  {author} {\bibfnamefont {S.~M.}\ \bibnamefont {Wiggins}}, \bibinfo {author}
  {\bibfnamefont {J.}~\bibnamefont {Rodier}}, \bibinfo {author} {\bibfnamefont
  {T.}~\bibnamefont {Garvey}},\ and\ \bibinfo {author} {\bibfnamefont {D.~A.}\
  \bibnamefont {Jaroszynski}},\ }\bibfield  {title} {\bibinfo {title} {Radial
  bunch compression: Path-length compensation in an rf photoinjector with a
  curved cathode},\ }\href {https://doi.org/10.1103/PhysRevSTAB.9.084201}
  {\bibfield  {journal} {\bibinfo  {journal} {Phys. Rev. ST Accel. Beams}\
  }\textbf {\bibinfo {volume} {9}},\ \bibinfo {pages} {084201} (\bibinfo {year}
  {2006})}\BibitemShut {NoStop}%
\bibitem [{\citenamefont {Cathey}\ \emph {et~al.}(2018)\citenamefont {Cathey},
  \citenamefont {Cousineau}, \citenamefont {Aleksandrov},\ and\ \citenamefont
  {Zhukov}}]{PhysRevLett.121.064804}%
  \BibitemOpen
  \bibfield  {author} {\bibinfo {author} {\bibfnamefont {B.}~\bibnamefont
  {Cathey}}, \bibinfo {author} {\bibfnamefont {S.}~\bibnamefont {Cousineau}},
  \bibinfo {author} {\bibfnamefont {A.}~\bibnamefont {Aleksandrov}},\ and\
  \bibinfo {author} {\bibfnamefont {A.}~\bibnamefont {Zhukov}},\ }\bibfield
  {title} {\bibinfo {title} {First six dimensional phase space measurement of
  an accelerator beam},\ }\href
  {https://doi.org/10.1103/PhysRevLett.121.064804} {\bibfield  {journal}
  {\bibinfo  {journal} {Phys. Rev. Lett.}\ }\textbf {\bibinfo {volume} {121}},\
  \bibinfo {pages} {064804} (\bibinfo {year} {2018})}\BibitemShut {NoStop}%
\bibitem [{\citenamefont {McKee}\ \emph {et~al.}(1995)\citenamefont {McKee},
  \citenamefont {O'Shea},\ and\ \citenamefont {Madey}}]{MCKEE1995264}%
  \BibitemOpen
  \bibfield  {author} {\bibinfo {author} {\bibfnamefont {C.}~\bibnamefont
  {McKee}}, \bibinfo {author} {\bibfnamefont {P.}~\bibnamefont {O'Shea}},\ and\
  \bibinfo {author} {\bibfnamefont {J.}~\bibnamefont {Madey}},\ }\bibfield
  {title} {\bibinfo {title} {Phase space tomography of relativistic electron
  beams},\ }\href {https://doi.org/10.1016/0168-9002(94)01411-6} {\bibfield
  {journal} {\bibinfo  {journal} {Nucl. Instrum. Methods Phys. Res., Sect. A}\
  }\textbf {\bibinfo {volume} {358}},\ \bibinfo {pages} {264} (\bibinfo {year}
  {1995})}\BibitemShut {NoStop}%
\bibitem [{\citenamefont {Stratakis}\ \emph {et~al.}(2006)\citenamefont
  {Stratakis}, \citenamefont {Kishek}, \citenamefont {Li}, \citenamefont
  {Bernal}, \citenamefont {Walter}, \citenamefont {Quinn}, \citenamefont
  {Reiser},\ and\ \citenamefont {O'Shea}}]{PhysRevSTAB.9.112801}%
  \BibitemOpen
  \bibfield  {author} {\bibinfo {author} {\bibfnamefont {D.}~\bibnamefont
  {Stratakis}}, \bibinfo {author} {\bibfnamefont {R.~A.}\ \bibnamefont
  {Kishek}}, \bibinfo {author} {\bibfnamefont {H.}~\bibnamefont {Li}}, \bibinfo
  {author} {\bibfnamefont {S.}~\bibnamefont {Bernal}}, \bibinfo {author}
  {\bibfnamefont {M.}~\bibnamefont {Walter}}, \bibinfo {author} {\bibfnamefont
  {B.}~\bibnamefont {Quinn}}, \bibinfo {author} {\bibfnamefont
  {M.}~\bibnamefont {Reiser}},\ and\ \bibinfo {author} {\bibfnamefont {P.~G.}\
  \bibnamefont {O'Shea}},\ }\bibfield  {title} {\bibinfo {title} {Tomography as
  a diagnostic tool for phase space mapping of intense particle beams},\ }\href
  {https://doi.org/10.1103/PhysRevSTAB.9.112801} {\bibfield  {journal}
  {\bibinfo  {journal} {Phys. Rev. Accel. Beams}\ }\textbf {\bibinfo {volume}
  {9}},\ \bibinfo {pages} {112801} (\bibinfo {year} {2006})}\BibitemShut
  {NoStop}%
\bibitem [{\citenamefont {Yakimenko}\ \emph {et~al.}(2003)\citenamefont
  {Yakimenko}, \citenamefont {Babzien}, \citenamefont {Ben-Zvi}, \citenamefont
  {Malone},\ and\ \citenamefont {Wang}}]{PhysRevSTAB.6.122801}%
  \BibitemOpen
  \bibfield  {author} {\bibinfo {author} {\bibfnamefont {V.}~\bibnamefont
  {Yakimenko}}, \bibinfo {author} {\bibfnamefont {M.}~\bibnamefont {Babzien}},
  \bibinfo {author} {\bibfnamefont {I.}~\bibnamefont {Ben-Zvi}}, \bibinfo
  {author} {\bibfnamefont {R.}~\bibnamefont {Malone}},\ and\ \bibinfo {author}
  {\bibfnamefont {X.-J.}\ \bibnamefont {Wang}},\ }\bibfield  {title} {\bibinfo
  {title} {Electron beam phase-space measurement using a high-precision
  tomography technique},\ }\href {https://doi.org/10.1103/PhysRevSTAB.6.122801}
  {\bibfield  {journal} {\bibinfo  {journal} {Phys. Rev. Accel. Beams}\
  }\textbf {\bibinfo {volume} {6}},\ \bibinfo {pages} {122801} (\bibinfo {year}
  {2003})}\BibitemShut {NoStop}%
\bibitem [{\citenamefont {Hermann}\ \emph {et~al.}(2021)\citenamefont
  {Hermann}, \citenamefont {Guzenko}, \citenamefont {H\"urzeler}, \citenamefont
  {Kirchner}, \citenamefont {Orlandi}, \citenamefont {Prat},\ and\
  \citenamefont {Ischebeck}}]{PhysRevAccelBeams.24.022802}%
  \BibitemOpen
  \bibfield  {author} {\bibinfo {author} {\bibfnamefont {B.}~\bibnamefont
  {Hermann}}, \bibinfo {author} {\bibfnamefont {V.~A.}\ \bibnamefont
  {Guzenko}}, \bibinfo {author} {\bibfnamefont {O.~R.}\ \bibnamefont
  {H\"urzeler}}, \bibinfo {author} {\bibfnamefont {A.}~\bibnamefont
  {Kirchner}}, \bibinfo {author} {\bibfnamefont {G.~L.}\ \bibnamefont
  {Orlandi}}, \bibinfo {author} {\bibfnamefont {E.}~\bibnamefont {Prat}},\ and\
  \bibinfo {author} {\bibfnamefont {R.}~\bibnamefont {Ischebeck}},\ }\bibfield
  {title} {\bibinfo {title} {Electron beam transverse phase space tomography
  using nanofabricated wire scanners with submicrometer resolution},\ }\href
  {https://doi.org/10.1103/PhysRevAccelBeams.24.022802} {\bibfield  {journal}
  {\bibinfo  {journal} {Phys. Rev. Accel. Beams}\ }\textbf {\bibinfo {volume}
  {24}},\ \bibinfo {pages} {022802} (\bibinfo {year} {2021})}\BibitemShut
  {NoStop}%
\bibitem [{\citenamefont {R\"ohrs}\ \emph {et~al.}(2009)\citenamefont
  {R\"ohrs}, \citenamefont {Gerth}, \citenamefont {Schlarb}, \citenamefont
  {Schmidt},\ and\ \citenamefont {Schm\"user}}]{PhysRevSTAB.12.050704}%
  \BibitemOpen
  \bibfield  {author} {\bibinfo {author} {\bibfnamefont {M.}~\bibnamefont
  {R\"ohrs}}, \bibinfo {author} {\bibfnamefont {C.}~\bibnamefont {Gerth}},
  \bibinfo {author} {\bibfnamefont {H.}~\bibnamefont {Schlarb}}, \bibinfo
  {author} {\bibfnamefont {B.}~\bibnamefont {Schmidt}},\ and\ \bibinfo {author}
  {\bibfnamefont {P.}~\bibnamefont {Schm\"user}},\ }\bibfield  {title}
  {\bibinfo {title} {Time-resolved electron beam phase space tomography at a
  soft x-ray free-electron laser},\ }\href
  {https://doi.org/10.1103/PhysRevSTAB.12.050704} {\bibfield  {journal}
  {\bibinfo  {journal} {Phys. Rev. Accel. Beams}\ }\textbf {\bibinfo {volume}
  {12}},\ \bibinfo {pages} {050704} (\bibinfo {year} {2009})}\BibitemShut
  {NoStop}%
\bibitem [{\citenamefont {Hock}\ and\ \citenamefont
  {Wolski}(2013)}]{HOCK20138}%
  \BibitemOpen
  \bibfield  {author} {\bibinfo {author} {\bibfnamefont {K.}~\bibnamefont
  {Hock}}\ and\ \bibinfo {author} {\bibfnamefont {A.}~\bibnamefont {Wolski}},\
  }\bibfield  {title} {\bibinfo {title} {Tomographic reconstruction of the full
  4{D} transverse phase space},\ }\href
  {https://doi.org/10.1016/j.nima.2013.05.004} {\bibfield  {journal} {\bibinfo
  {journal} {Nucl. Instrum. Methods Phys. Res., Sect. A}\ }\textbf {\bibinfo
  {volume} {726}},\ \bibinfo {pages} {8} (\bibinfo {year} {2013})}\BibitemShut
  {NoStop}%
\bibitem [{\citenamefont {Wolski}\ \emph {et~al.}(2020)\citenamefont {Wolski},
  \citenamefont {Christie}, \citenamefont {Militsyn}, \citenamefont {Scott},\
  and\ \citenamefont {Kockelbergh}}]{PhysRevAccelBeams.23.032804}%
  \BibitemOpen
  \bibfield  {author} {\bibinfo {author} {\bibfnamefont {A.}~\bibnamefont
  {Wolski}}, \bibinfo {author} {\bibfnamefont {D.~C.}\ \bibnamefont
  {Christie}}, \bibinfo {author} {\bibfnamefont {B.~L.}\ \bibnamefont
  {Militsyn}}, \bibinfo {author} {\bibfnamefont {D.~J.}\ \bibnamefont
  {Scott}},\ and\ \bibinfo {author} {\bibfnamefont {H.}~\bibnamefont
  {Kockelbergh}},\ }\bibfield  {title} {\bibinfo {title} {Transverse phase
  space characterization in an accelerator test facility},\ }\href
  {https://doi.org/10.1103/PhysRevAccelBeams.23.032804} {\bibfield  {journal}
  {\bibinfo  {journal} {Phys. Rev. Accel. Beams}\ }\textbf {\bibinfo {volume}
  {23}},\ \bibinfo {pages} {032804} (\bibinfo {year} {2020})}\BibitemShut
  {NoStop}%
\bibitem [{\citenamefont {Guo}\ \emph {et~al.}(2021)\citenamefont {Guo},
  \citenamefont {Denham}, \citenamefont {Musumeci}, \citenamefont {Ody},\ and\
  \citenamefont {Park}}]{guo:ibic21-tupp15}%
  \BibitemOpen
  \bibfield  {author} {\bibinfo {author} {\bibfnamefont {V.}~\bibnamefont
  {Guo}}, \bibinfo {author} {\bibfnamefont {P.~E.}\ \bibnamefont {Denham}},
  \bibinfo {author} {\bibfnamefont {P.}~\bibnamefont {Musumeci}}, \bibinfo
  {author} {\bibfnamefont {A.}~\bibnamefont {Ody}},\ and\ \bibinfo {author}
  {\bibfnamefont {Y.}~\bibnamefont {Park}},\ }\bibfield  {title} {\bibinfo
  {title} {{4D Beam Tomography at the UCLA Pegasus Laboratory}},\ }in\ \href
  {https://doi.org/10.18429/JACoW-IBIC2021-TUPP15} {\emph {\bibinfo {booktitle}
  {Proc. IBIC'21}}}\ (\bibinfo  {publisher} {JACoW Publishing, Geneva,
  Switzerland},\ \bibinfo {year} {2021})\BibitemShut {NoStop}%
\bibitem [{\citenamefont {Jaster-Merz}\ \emph {et~al.}(2021)\citenamefont
  {Jaster-Merz}, \citenamefont {Assmann}, \citenamefont {Brinkmann},
  \citenamefont {Burkart}, \citenamefont {Dinter}, \citenamefont {Mayet},\ and\
  \citenamefont {Kuropka}}]{jaster-merz:ipac21-mopab302}%
  \BibitemOpen
  \bibfield  {author} {\bibinfo {author} {\bibfnamefont {S.}~\bibnamefont
  {Jaster-Merz}}, \bibinfo {author} {\bibfnamefont {R.}~\bibnamefont
  {Assmann}}, \bibinfo {author} {\bibfnamefont {R.}~\bibnamefont {Brinkmann}},
  \bibinfo {author} {\bibfnamefont {F.}~\bibnamefont {Burkart}}, \bibinfo
  {author} {\bibfnamefont {H.}~\bibnamefont {Dinter}}, \bibinfo {author}
  {\bibfnamefont {F.}~\bibnamefont {Mayet}},\ and\ \bibinfo {author}
  {\bibfnamefont {W.}~\bibnamefont {Kuropka}},\ }\bibfield  {title} {\bibinfo
  {title} {Characterization of the full transverse phase space of electron
  bunches at ares},\ }in\ \href
  {https://doi.org/10.18429/JACoW-IPAC2021-MOPAB302} {\emph {\bibinfo
  {booktitle} {12th Int. Particle Accelerator Conf.(IPAC’21), Campinas,
  Brazil}}}\ (\bibinfo {year} {2021})\BibitemShut {NoStop}%
\bibitem [{\citenamefont {Marx}\ \emph
  {et~al.}(2017{\natexlab{a}})\citenamefont {Marx}, \citenamefont {Assmann},
  \citenamefont {Craievich}, \citenamefont {Dorda}, \citenamefont {Grudiev},\
  and\ \citenamefont {Marchetti}}]{Marx_2017}%
  \BibitemOpen
  \bibfield  {author} {\bibinfo {author} {\bibfnamefont {D.}~\bibnamefont
  {Marx}}, \bibinfo {author} {\bibfnamefont {R.}~\bibnamefont {Assmann}},
  \bibinfo {author} {\bibfnamefont {P.}~\bibnamefont {Craievich}}, \bibinfo
  {author} {\bibfnamefont {U.}~\bibnamefont {Dorda}}, \bibinfo {author}
  {\bibfnamefont {A.}~\bibnamefont {Grudiev}},\ and\ \bibinfo {author}
  {\bibfnamefont {B.}~\bibnamefont {Marchetti}},\ }\bibfield  {title} {\bibinfo
  {title} {Reconstruction of the 3{D} charge distribution of an electron bunch
  using a novel variable-polarization transverse deflecting structure
  ({TDS})},\ }\href {https://doi.org/10.1088/1742-6596/874/1/012077} {\bibfield
   {journal} {\bibinfo  {journal} {Journal of Physics: Conference Series}\
  }\textbf {\bibinfo {volume} {874}},\ \bibinfo {pages} {012077} (\bibinfo
  {year} {2017}{\natexlab{a}})}\BibitemShut {NoStop}%
\bibitem [{\citenamefont {Marx}\ \emph {et~al.}(2019)\citenamefont {Marx},
  \citenamefont {Assmann}, \citenamefont {Craievich}, \citenamefont
  {Floettmann}, \citenamefont {Grudiev},\ and\ \citenamefont
  {Marchetti}}]{Marx2019}%
  \BibitemOpen
  \bibfield  {author} {\bibinfo {author} {\bibfnamefont {D.}~\bibnamefont
  {Marx}}, \bibinfo {author} {\bibfnamefont {R.~W.}\ \bibnamefont {Assmann}},
  \bibinfo {author} {\bibfnamefont {P.}~\bibnamefont {Craievich}}, \bibinfo
  {author} {\bibfnamefont {K.}~\bibnamefont {Floettmann}}, \bibinfo {author}
  {\bibfnamefont {A.}~\bibnamefont {Grudiev}},\ and\ \bibinfo {author}
  {\bibfnamefont {B.}~\bibnamefont {Marchetti}},\ }\bibfield  {title} {\bibinfo
  {title} {Simulation studies for characterizing ultrashort bunches using novel
  polarizable {X}-band transverse deflection structures},\ }\href
  {https://doi.org/10.1038/s41598-019-56433-8} {\bibfield  {journal} {\bibinfo
  {journal} {Sci. Rep.}\ }\textbf {\bibinfo {volume} {9}},\ \bibinfo {pages}
  {19912} (\bibinfo {year} {2019})}\BibitemShut {NoStop}%
\bibitem [{\citenamefont {Marx}(2019)}]{Marx:427780}%
  \BibitemOpen
  \bibfield  {author} {\bibinfo {author} {\bibfnamefont {D.}~\bibnamefont
  {Marx}},\ }\emph {\bibinfo {title} {{C}haracterization of {U}ltrashort
  {E}lectron {B}unches at the {SINBAD}-{ARES} {L}inac}},\ \href
  {https://doi.org/10.3204/PUBDB-2019-04190} {\bibinfo {type} {Dissertation}},\
  \bibinfo  {school} {Universität Hamburg}, \bibinfo {address} {Hamburg}
  (\bibinfo {year} {2019}),\ \bibinfo {note} {dissertation, Universität
  Hamburg, 2019}\BibitemShut {NoStop}%
\bibitem [{\citenamefont {Marchetti}\ \emph {et~al.}(2021)\citenamefont
  {Marchetti} \emph {et~al.}}]{Marchetti2021}%
  \BibitemOpen
  \bibfield  {author} {\bibinfo {author} {\bibfnamefont {B.}~\bibnamefont
  {Marchetti}} \emph {et~al.},\ }\bibfield  {title} {\bibinfo {title}
  {Experimental demonstration of novel beam characterization using a
  polarizable {X}-band transverse deflection structure},\ }\href
  {https://doi.org/10.1038/s41598-021-82687-2} {\bibfield  {journal} {\bibinfo
  {journal} {Sci. Rep.}\ }\textbf {\bibinfo {volume} {11}},\ \bibinfo {pages}
  {3560} (\bibinfo {year} {2021})}\BibitemShut {NoStop}%
\bibitem [{\citenamefont {Scheinker}\ \emph {et~al.}(2018)\citenamefont
  {Scheinker}, \citenamefont {Edelen}, \citenamefont {Bohler}, \citenamefont
  {Emma},\ and\ \citenamefont {Lutman}}]{PhysRevLett.121.044801}%
  \BibitemOpen
  \bibfield  {author} {\bibinfo {author} {\bibfnamefont {A.}~\bibnamefont
  {Scheinker}}, \bibinfo {author} {\bibfnamefont {A.}~\bibnamefont {Edelen}},
  \bibinfo {author} {\bibfnamefont {D.}~\bibnamefont {Bohler}}, \bibinfo
  {author} {\bibfnamefont {C.}~\bibnamefont {Emma}},\ and\ \bibinfo {author}
  {\bibfnamefont {A.}~\bibnamefont {Lutman}},\ }\bibfield  {title} {\bibinfo
  {title} {Demonstration of model-independent control of the longitudinal phase
  space of electron beams in the linac-coherent light source with femtosecond
  resolution},\ }\href {https://doi.org/10.1103/PhysRevLett.121.044801}
  {\bibfield  {journal} {\bibinfo  {journal} {Phys. Rev. Lett.}\ }\textbf
  {\bibinfo {volume} {121}},\ \bibinfo {pages} {044801} (\bibinfo {year}
  {2018})}\BibitemShut {NoStop}%
\bibitem [{\citenamefont {Scheinker}\ \emph {et~al.}(2022)\citenamefont
  {Scheinker}, \citenamefont {Cropp~V},\ and\ \citenamefont
  {Filippetto}}]{Scheinker2022}%
  \BibitemOpen
  \bibfield  {author} {\bibinfo {author} {\bibfnamefont {A.}~\bibnamefont
  {Scheinker}}, \bibinfo {author} {\bibfnamefont {F.~W.}\ \bibnamefont
  {Cropp~V}},\ and\ \bibinfo {author} {\bibfnamefont {D.}~\bibnamefont
  {Filippetto}},\ }\bibfield  {title} {\bibinfo {title} {{6D Phase Space
  Diagnostics Based on Adaptively Tuned Physics-Informed Generative
  Convolutional Neural Networks}},\ }in\ \href
  {https://doi.org/10.18429/JACoW-IPAC2022-TUOXGD3} {\emph {\bibinfo
  {booktitle} {Proceedings of the 13th International Particle Accelerator
  Conference, Bangkok, Thailand}}}\ (\bibinfo {year} {2022})\ pp.\ \bibinfo
  {pages} {776--779}\BibitemShut {NoStop}%
\bibitem [{\citenamefont {Craievich}\ \emph {et~al.}(2020)\citenamefont
  {Craievich} \emph {et~al.}}]{PhysRevAccelBeams.23.112001}%
  \BibitemOpen
  \bibfield  {author} {\bibinfo {author} {\bibfnamefont {P.}~\bibnamefont
  {Craievich}} \emph {et~al.},\ }\bibfield  {title} {\bibinfo {title} {Novel
  {$X$}-band transverse deflection structure with variable polarization},\
  }\href {https://doi.org/10.1103/PhysRevAccelBeams.23.112001} {\bibfield
  {journal} {\bibinfo  {journal} {Phys. Rev. Accel. Beams}\ }\textbf {\bibinfo
  {volume} {23}},\ \bibinfo {pages} {112001} (\bibinfo {year}
  {2020})}\BibitemShut {NoStop}%
\bibitem [{\citenamefont {Grudiev}(2016)}]{Grudiev:2158484}%
  \BibitemOpen
  \bibfield  {author} {\bibinfo {author} {\bibfnamefont {A.}~\bibnamefont
  {Grudiev}},\ }\bibfield  {title} {\bibinfo {title} {{design of compact high
  power rf components at x-band}},\ }\href {https://cds.cern.ch/record/2158484}
  {\bibfield  {journal} {\bibinfo  {journal} {CLIC note}\ } (\bibinfo {year}
  {2016})}\BibitemShut {NoStop}%
\bibitem [{\citenamefont {Burkart}\ \emph {et~al.}(2022)\citenamefont
  {Burkart}, \citenamefont {Assmann}, \citenamefont {Dinter}, \citenamefont
  {Jaster-Merz}, \citenamefont {Kuropka}, \citenamefont {Mayet},\ and\
  \citenamefont {Vinatier}}]{burkart:linac2022-thpojo01}%
  \BibitemOpen
  \bibfield  {author} {\bibinfo {author} {\bibfnamefont {F.}~\bibnamefont
  {Burkart}}, \bibinfo {author} {\bibfnamefont {R.}~\bibnamefont {Assmann}},
  \bibinfo {author} {\bibfnamefont {H.}~\bibnamefont {Dinter}}, \bibinfo
  {author} {\bibfnamefont {S.}~\bibnamefont {Jaster-Merz}}, \bibinfo {author}
  {\bibfnamefont {W.}~\bibnamefont {Kuropka}}, \bibinfo {author} {\bibfnamefont
  {F.}~\bibnamefont {Mayet}},\ and\ \bibinfo {author} {\bibfnamefont
  {T.}~\bibnamefont {Vinatier}},\ }\bibfield  {title} {\bibinfo {title} {{The
  ARES Linac at DESY}},\ }in\ \href
  {https://doi.org/10.18429/JACoW-LINAC2022-THPOJO01} {\emph {\bibinfo
  {booktitle} {Proc. LINAC'22}}},\ \bibinfo {series and number} {\bibinfo
  {series} {International Linear Accelerator Conference}\ No.~\bibinfo {number}
  {31}}\ (\bibinfo  {publisher} {JACoW Publishing, Geneva, Switzerland},\
  \bibinfo {year} {2022})\ pp.\ \bibinfo {pages} {691--694}\BibitemShut
  {NoStop}%
\bibitem [{\citenamefont {Marchetti}\ \emph {et~al.}(2020)\citenamefont
  {Marchetti} \emph {et~al.}}]{Marchetti_2020}%
  \BibitemOpen
  \bibfield  {author} {\bibinfo {author} {\bibfnamefont {B.}~\bibnamefont
  {Marchetti}} \emph {et~al.},\ }\bibfield  {title} {\bibinfo {title}
  {{SINBAD}-{ARES} - a photo-injector for external injection experiments in
  novel accelerators at {DESY}},\ }\href
  {https://doi.org/10.1088/1742-6596/1596/1/012036} {\bibfield  {journal}
  {\bibinfo  {journal} {Journal of Physics: Conference Series}\ }\textbf
  {\bibinfo {volume} {1596}},\ \bibinfo {pages} {012036} (\bibinfo {year}
  {2020})}\BibitemShut {NoStop}%
\bibitem [{\citenamefont {Panofski}\ \emph {et~al.}(2021)\citenamefont
  {Panofski}, \citenamefont {Assmann}, \citenamefont {Burkart}, \citenamefont
  {Dorda}, \citenamefont {Genovese}, \citenamefont {Jafarinia}, \citenamefont
  {Jaster-Merz}, \citenamefont {Kellermeier}, \citenamefont {Kuropka},
  \citenamefont {Lemery}, \citenamefont {Marchetti}, \citenamefont {Marx},
  \citenamefont {Mayet}, \citenamefont {Vinatier},\ and\ \citenamefont
  {Yamin}}]{instruments5030028}%
  \BibitemOpen
  \bibfield  {author} {\bibinfo {author} {\bibfnamefont {E.}~\bibnamefont
  {Panofski}}, \bibinfo {author} {\bibfnamefont {R.}~\bibnamefont {Assmann}},
  \bibinfo {author} {\bibfnamefont {F.}~\bibnamefont {Burkart}}, \bibinfo
  {author} {\bibfnamefont {U.}~\bibnamefont {Dorda}}, \bibinfo {author}
  {\bibfnamefont {L.}~\bibnamefont {Genovese}}, \bibinfo {author}
  {\bibfnamefont {F.}~\bibnamefont {Jafarinia}}, \bibinfo {author}
  {\bibfnamefont {S.}~\bibnamefont {Jaster-Merz}}, \bibinfo {author}
  {\bibfnamefont {M.}~\bibnamefont {Kellermeier}}, \bibinfo {author}
  {\bibfnamefont {W.}~\bibnamefont {Kuropka}}, \bibinfo {author} {\bibfnamefont
  {F.}~\bibnamefont {Lemery}}, \bibinfo {author} {\bibfnamefont
  {B.}~\bibnamefont {Marchetti}}, \bibinfo {author} {\bibfnamefont
  {D.}~\bibnamefont {Marx}}, \bibinfo {author} {\bibfnamefont {F.}~\bibnamefont
  {Mayet}}, \bibinfo {author} {\bibfnamefont {T.}~\bibnamefont {Vinatier}},\
  and\ \bibinfo {author} {\bibfnamefont {S.}~\bibnamefont {Yamin}},\ }\bibfield
   {title} {\bibinfo {title} {Commissioning results and electron beam
  characterization with the {S}-band photoinjector at {SINBAD-ARES}},\
  }\bibfield  {journal} {\bibinfo  {journal} {Instruments}\ }\textbf {\bibinfo
  {volume} {5}},\ \href {https://doi.org/10.3390/instruments5030028}
  {10.3390/instruments5030028} (\bibinfo {year} {2021})\BibitemShut {NoStop}%
\bibitem [{\citenamefont {Dorda}\ \emph {et~al.}(2018)\citenamefont {Dorda},
  \citenamefont {Marchetti}, \citenamefont {Zhu}, \citenamefont {Mayet},
  \citenamefont {Kuropka}, \citenamefont {Vinatier}, \citenamefont
  {Vashchenko}, \citenamefont {Galaydych}, \citenamefont {Walker},
  \citenamefont {Marx}, \citenamefont {Brinkmann}, \citenamefont {Assmann},
  \citenamefont {Matlis}, \citenamefont {Fallahi},\ and\ \citenamefont
  {Kaertner}}]{DORDA2018239}%
  \BibitemOpen
  \bibfield  {author} {\bibinfo {author} {\bibfnamefont {U.}~\bibnamefont
  {Dorda}}, \bibinfo {author} {\bibfnamefont {B.}~\bibnamefont {Marchetti}},
  \bibinfo {author} {\bibfnamefont {J.}~\bibnamefont {Zhu}}, \bibinfo {author}
  {\bibfnamefont {F.}~\bibnamefont {Mayet}}, \bibinfo {author} {\bibfnamefont
  {W.}~\bibnamefont {Kuropka}}, \bibinfo {author} {\bibfnamefont
  {T.}~\bibnamefont {Vinatier}}, \bibinfo {author} {\bibfnamefont
  {G.}~\bibnamefont {Vashchenko}}, \bibinfo {author} {\bibfnamefont
  {K.}~\bibnamefont {Galaydych}}, \bibinfo {author} {\bibfnamefont
  {P.}~\bibnamefont {Walker}}, \bibinfo {author} {\bibfnamefont
  {D.}~\bibnamefont {Marx}}, \bibinfo {author} {\bibfnamefont {R.}~\bibnamefont
  {Brinkmann}}, \bibinfo {author} {\bibfnamefont {R.}~\bibnamefont {Assmann}},
  \bibinfo {author} {\bibfnamefont {N.}~\bibnamefont {Matlis}}, \bibinfo
  {author} {\bibfnamefont {A.}~\bibnamefont {Fallahi}},\ and\ \bibinfo {author}
  {\bibfnamefont {F.}~\bibnamefont {Kaertner}},\ }\bibfield  {title} {\bibinfo
  {title} {Status and objectives of the dedicated accelerator {R}\&{D} facility
  “{SINBAD}” at {DESY}},\ }\href
  {https://doi.org/10.1016/j.nima.2018.01.036} {\bibfield  {journal} {\bibinfo
  {journal} {Nucl. Instrum. Methods Phys. Res., Sect. A}\ }\textbf {\bibinfo
  {volume} {909}},\ \bibinfo {pages} {239} (\bibinfo {year} {2018})},\ \bibinfo
  {note} {3rd European Advanced Accelerator Concepts workshop
  (EAAC2017)}\BibitemShut {NoStop}%
\bibitem [{\citenamefont {Jaster-Merz}\ \emph
  {et~al.}(2022{\natexlab{a}})\citenamefont {Jaster-Merz}, \citenamefont
  {Assmann}, \citenamefont {Brinkmann}, \citenamefont {Burkart},\ and\
  \citenamefont {Vinatier}}]{jaster-merz:ipac22-mopopt021}%
  \BibitemOpen
  \bibfield  {author} {\bibinfo {author} {\bibfnamefont {S.}~\bibnamefont
  {Jaster-Merz}}, \bibinfo {author} {\bibfnamefont {R.}~\bibnamefont
  {Assmann}}, \bibinfo {author} {\bibfnamefont {R.}~\bibnamefont {Brinkmann}},
  \bibinfo {author} {\bibfnamefont {F.}~\bibnamefont {Burkart}},\ and\ \bibinfo
  {author} {\bibfnamefont {T.}~\bibnamefont {Vinatier}},\ }\bibfield  {title}
  {\bibinfo {title} {{5D tomography of electron bunches at ARES}},\ }in\ \href
  {https://doi.org/10.18429/JACoW-IPAC2022-MOPOPT021} {\emph {\bibinfo
  {booktitle} {Proceedings of the 13th International Particle Accelerator
  Conference, Bangkok, Thailand}}}\ (\bibinfo {year} {2022})\ pp.\ \bibinfo
  {pages} {279--283}\BibitemShut {NoStop}%
\bibitem [{\citenamefont {Jaster-Merz}\ \emph
  {et~al.}(2022{\natexlab{b}})\citenamefont {Jaster-Merz}, \citenamefont
  {Aßmann}, \citenamefont {Brinkmann}, \citenamefont {Burkart},\ and\
  \citenamefont {Vinatier}}]{jaster-merz:linac2022-mopori10}%
  \BibitemOpen
  \bibfield  {author} {\bibinfo {author} {\bibfnamefont {S.}~\bibnamefont
  {Jaster-Merz}}, \bibinfo {author} {\bibfnamefont {R.}~\bibnamefont
  {Aßmann}}, \bibinfo {author} {\bibfnamefont {R.}~\bibnamefont {Brinkmann}},
  \bibinfo {author} {\bibfnamefont {F.}~\bibnamefont {Burkart}},\ and\ \bibinfo
  {author} {\bibfnamefont {T.}~\bibnamefont {Vinatier}},\ }\bibfield  {title}
  {\bibinfo {title} {{First Studies of 5D Phase-Space Tomography of Electron
  Beams at ARES}},\ }in\ \href
  {https://doi.org/10.18429/JACoW-LINAC2022-MOPORI10} {\emph {\bibinfo
  {booktitle} {Proc. LINAC'22}}},\ \bibinfo {series and number} {\bibinfo
  {series} {International Linear Accelerator Conference}\ No.~\bibinfo {number}
  {31}}\ (\bibinfo  {publisher} {JACoW Publishing, Geneva, Switzerland},\
  \bibinfo {year} {2022})\ pp.\ \bibinfo {pages} {247--251}\BibitemShut
  {NoStop}%
\bibitem [{\citenamefont {Kak}\ and\ \citenamefont
  {Slaney}(2001)}]{kak2001principles}%
  \BibitemOpen
  \bibfield  {author} {\bibinfo {author} {\bibfnamefont {A.~C.}\ \bibnamefont
  {Kak}}\ and\ \bibinfo {author} {\bibfnamefont {M.}~\bibnamefont {Slaney}},\
  }\href@noop {} {\emph {\bibinfo {title} {Principles of computerized
  tomographic imaging}}}\ (\bibinfo  {publisher} {SIAM},\ \bibinfo {year}
  {2001})\BibitemShut {NoStop}%
\bibitem [{\citenamefont {Gordon}\ \emph {et~al.}(1970)\citenamefont {Gordon},
  \citenamefont {Bender},\ and\ \citenamefont {Herman}}]{GORDON1970471}%
  \BibitemOpen
  \bibfield  {author} {\bibinfo {author} {\bibfnamefont {R.}~\bibnamefont
  {Gordon}}, \bibinfo {author} {\bibfnamefont {R.}~\bibnamefont {Bender}},\
  and\ \bibinfo {author} {\bibfnamefont {G.~T.}\ \bibnamefont {Herman}},\
  }\bibfield  {title} {\bibinfo {title} {Algebraic reconstruction techniques
  (art) for three-dimensional electron microscopy and x-ray photography},\
  }\href {https://doi.org/https://doi.org/10.1016/0022-5193(70)90109-8}
  {\bibfield  {journal} {\bibinfo  {journal} {Journal of Theoretical Biology}\
  }\textbf {\bibinfo {volume} {29}},\ \bibinfo {pages} {471} (\bibinfo {year}
  {1970})}\BibitemShut {NoStop}%
\bibitem [{\citenamefont {Andersen}\ and\ \citenamefont
  {Kak}(1984)}]{ANDERSEN198481}%
  \BibitemOpen
  \bibfield  {author} {\bibinfo {author} {\bibfnamefont {A.}~\bibnamefont
  {Andersen}}\ and\ \bibinfo {author} {\bibfnamefont {A.}~\bibnamefont {Kak}},\
  }\bibfield  {title} {\bibinfo {title} {{S}imultaneous {A}lgebraic
  {R}econstruction {T}echnique ({SART}): A superior implementation of the {ART}
  algorithm},\ }\href {https://doi.org/10.1016/0161-7346(84)90008-7} {\bibfield
   {journal} {\bibinfo  {journal} {Ultrason. Imaging}\ }\textbf {\bibinfo
  {volume} {6}},\ \bibinfo {pages} {81} (\bibinfo {year} {1984})}\BibitemShut
  {NoStop}%
\bibitem [{\citenamefont {Minerbo}(1979)}]{MINERBO197948}%
  \BibitemOpen
  \bibfield  {author} {\bibinfo {author} {\bibfnamefont {G.}~\bibnamefont
  {Minerbo}},\ }\bibfield  {title} {\bibinfo {title} {Ment: A maximum entropy
  algorithm for reconstructing a source from projection data},\ }\href
  {https://doi.org/https://doi.org/10.1016/0146-664X(79)90034-0} {\bibfield
  {journal} {\bibinfo  {journal} {Computer Graphics and Image Processing}\
  }\textbf {\bibinfo {volume} {10}},\ \bibinfo {pages} {48} (\bibinfo {year}
  {1979})}\BibitemShut {NoStop}%
\bibitem [{\citenamefont {Hock}\ \emph {et~al.}(2011)\citenamefont {Hock},
  \citenamefont {Ibison}, \citenamefont {Holder}, \citenamefont {Wolski},\ and\
  \citenamefont {Muratori}}]{HOCK201136}%
  \BibitemOpen
  \bibfield  {author} {\bibinfo {author} {\bibfnamefont {K.}~\bibnamefont
  {Hock}}, \bibinfo {author} {\bibfnamefont {M.}~\bibnamefont {Ibison}},
  \bibinfo {author} {\bibfnamefont {D.}~\bibnamefont {Holder}}, \bibinfo
  {author} {\bibfnamefont {A.}~\bibnamefont {Wolski}},\ and\ \bibinfo {author}
  {\bibfnamefont {B.}~\bibnamefont {Muratori}},\ }\bibfield  {title} {\bibinfo
  {title} {Beam tomography in transverse normalised phase space},\ }\href
  {https://doi.org/10.1016/j.nima.2011.04.002} {\bibfield  {journal} {\bibinfo
  {journal} {Nucl. Instrum. Methods Phys. Res., Sect. A}\ }\textbf {\bibinfo
  {volume} {642}},\ \bibinfo {pages} {36} (\bibinfo {year} {2011})}\BibitemShut
  {NoStop}%
\bibitem [{\citenamefont {Courant}\ and\ \citenamefont
  {Snyder}(1958)}]{COURANT19581}%
  \BibitemOpen
  \bibfield  {author} {\bibinfo {author} {\bibfnamefont {E.}~\bibnamefont
  {Courant}}\ and\ \bibinfo {author} {\bibfnamefont {H.}~\bibnamefont
  {Snyder}},\ }\bibfield  {title} {\bibinfo {title} {Theory of the
  alternating-gradient synchrotron},\ }\href
  {https://doi.org/10.1016/0003-4916(58)90012-5} {\bibfield  {journal}
  {\bibinfo  {journal} {Annals of Physics}\ }\textbf {\bibinfo {volume} {3}},\
  \bibinfo {pages} {1} (\bibinfo {year} {1958})}\BibitemShut {NoStop}%
\bibitem [{\citenamefont {Marx}\ \emph
  {et~al.}(2017{\natexlab{b}})\citenamefont {Marx}, \citenamefont {Assmann},
  \citenamefont {Dorda}, \citenamefont {Marchetti},\ and\ \citenamefont
  {Mayet}}]{Marx_2017_TDS}%
  \BibitemOpen
  \bibfield  {author} {\bibinfo {author} {\bibfnamefont {D.}~\bibnamefont
  {Marx}}, \bibinfo {author} {\bibfnamefont {R.}~\bibnamefont {Assmann}},
  \bibinfo {author} {\bibfnamefont {U.}~\bibnamefont {Dorda}}, \bibinfo
  {author} {\bibfnamefont {B.}~\bibnamefont {Marchetti}},\ and\ \bibinfo
  {author} {\bibfnamefont {F.}~\bibnamefont {Mayet}},\ }\bibfield  {title}
  {\bibinfo {title} {Lattice considerations for the use of an {X}-band
  transverse deflecting structure ({TDS}) at {SINBAD}, {DESY}},\ }\href
  {https://doi.org/10.1088/1742-6596/874/1/012078} {\bibfield  {journal}
  {\bibinfo  {journal} {Journal of Physics: Conference Series}\ }\textbf
  {\bibinfo {volume} {874}},\ \bibinfo {pages} {012078} (\bibinfo {year}
  {2017}{\natexlab{b}})}\BibitemShut {NoStop}%
\bibitem [{\citenamefont {Agapov}\ \emph {et~al.}(2014)\citenamefont {Agapov},
  \citenamefont {Geloni}, \citenamefont {Tomin},\ and\ \citenamefont
  {Zagorodnov}}]{AGAPOV2014151}%
  \BibitemOpen
  \bibfield  {author} {\bibinfo {author} {\bibfnamefont {I.}~\bibnamefont
  {Agapov}}, \bibinfo {author} {\bibfnamefont {G.}~\bibnamefont {Geloni}},
  \bibinfo {author} {\bibfnamefont {S.}~\bibnamefont {Tomin}},\ and\ \bibinfo
  {author} {\bibfnamefont {I.}~\bibnamefont {Zagorodnov}},\ }\bibfield  {title}
  {\bibinfo {title} {{OCELOT}: A software framework for synchrotron light
  source and {FEL} studies},\ }\href
  {https://doi.org/10.1016/j.nima.2014.09.057} {\bibfield  {journal} {\bibinfo
  {journal} {Nucl. Instrum. Methods Phys. Res., Sect. A}\ }\textbf {\bibinfo
  {volume} {768}},\ \bibinfo {pages} {151} (\bibinfo {year}
  {2014})}\BibitemShut {NoStop}%
\bibitem [{\citenamefont {Tomin}\ \emph {et~al.}(2017)\citenamefont {Tomin},
  \citenamefont {Agapov}, \citenamefont {Dohlus},\ and\ \citenamefont
  {Zagorodnov}}]{tomin:ipac17-wepab031}%
  \BibitemOpen
  \bibfield  {author} {\bibinfo {author} {\bibfnamefont {S.~I.}\ \bibnamefont
  {Tomin}}, \bibinfo {author} {\bibfnamefont {I.~V.}\ \bibnamefont {Agapov}},
  \bibinfo {author} {\bibfnamefont {M.}~\bibnamefont {Dohlus}},\ and\ \bibinfo
  {author} {\bibfnamefont {I.}~\bibnamefont {Zagorodnov}},\ }\bibfield  {title}
  {\bibinfo {title} {{OCELOT as a Framework for Beam Dynamics Simulations of
  X-Ray Sources}},\ }in\ \href
  {https://doi.org/10.18429/JACoW-IPAC2017-WEPAB031} {\emph {\bibinfo
  {booktitle} {Proc. IPAC'17}}}\ (\bibinfo  {publisher} {JACoW Publishing,
  Geneva, Switzerland},\ \bibinfo {year} {2017})\BibitemShut {NoStop}%
\bibitem [{\citenamefont {Anderson}\ \emph {et~al.}(2002)\citenamefont
  {Anderson}, \citenamefont {Rosenzweig}, \citenamefont {LeSage},\ and\
  \citenamefont {Crane}}]{PhysRevSTAB.5.014201}%
  \BibitemOpen
  \bibfield  {author} {\bibinfo {author} {\bibfnamefont {S.~G.}\ \bibnamefont
  {Anderson}}, \bibinfo {author} {\bibfnamefont {J.~B.}\ \bibnamefont
  {Rosenzweig}}, \bibinfo {author} {\bibfnamefont {G.~P.}\ \bibnamefont
  {LeSage}},\ and\ \bibinfo {author} {\bibfnamefont {J.~K.}\ \bibnamefont
  {Crane}},\ }\bibfield  {title} {\bibinfo {title} {Space-charge effects in
  high brightness electron beam emittance measurements},\ }\href
  {https://doi.org/10.1103/PhysRevSTAB.5.014201} {\bibfield  {journal}
  {\bibinfo  {journal} {Phys. Rev. ST Accel. Beams}\ }\textbf {\bibinfo
  {volume} {5}},\ \bibinfo {pages} {014201} (\bibinfo {year}
  {2002})}\BibitemShut {NoStop}%
\bibitem [{\citenamefont {Van~der Walt}\ \emph {et~al.}(2014)\citenamefont
  {Van~der Walt}, \citenamefont {Sch{\"o}nberger}, \citenamefont
  {Nunez-Iglesias}, \citenamefont {Boulogne}, \citenamefont {Warner},
  \citenamefont {Yager}, \citenamefont {Gouillart},\ and\ \citenamefont
  {Yu}}]{van2014scikit}%
  \BibitemOpen
  \bibfield  {author} {\bibinfo {author} {\bibfnamefont {S.}~\bibnamefont
  {Van~der Walt}}, \bibinfo {author} {\bibfnamefont {J.~L.}\ \bibnamefont
  {Sch{\"o}nberger}}, \bibinfo {author} {\bibfnamefont {J.}~\bibnamefont
  {Nunez-Iglesias}}, \bibinfo {author} {\bibfnamefont {F.}~\bibnamefont
  {Boulogne}}, \bibinfo {author} {\bibfnamefont {J.~D.}\ \bibnamefont
  {Warner}}, \bibinfo {author} {\bibfnamefont {N.}~\bibnamefont {Yager}},
  \bibinfo {author} {\bibfnamefont {E.}~\bibnamefont {Gouillart}},\ and\
  \bibinfo {author} {\bibfnamefont {T.}~\bibnamefont {Yu}},\ }\bibfield
  {title} {\bibinfo {title} {scikit-image: image processing in {P}ython},\
  }\href {https://doi.org/10.7717/peerj.453} {\bibfield  {journal} {\bibinfo
  {journal} {PeerJ}\ }\textbf {\bibinfo {volume} {2}},\ \bibinfo {pages} {e453}
  (\bibinfo {year} {2014})}\BibitemShut {NoStop}%
\bibitem [{\citenamefont {Sullivan}\ and\ \citenamefont
  {Kaszynski}(2019)}]{sullivan2019pyvista}%
  \BibitemOpen
  \bibfield  {author} {\bibinfo {author} {\bibfnamefont {B.}~\bibnamefont
  {Sullivan}}\ and\ \bibinfo {author} {\bibfnamefont {A.}~\bibnamefont
  {Kaszynski}},\ }\bibfield  {title} {\bibinfo {title} {{PyVista}: {3D}
  plotting and mesh analysis through a streamlined interface for the
  {Visualization Toolkit} ({VTK})},\ }\href
  {https://doi.org/10.21105/joss.01450} {\bibfield  {journal} {\bibinfo
  {journal} {Journal of Open Source Software}\ }\textbf {\bibinfo {volume}
  {4}},\ \bibinfo {pages} {1450} (\bibinfo {year} {2019})}\BibitemShut
  {NoStop}%
\bibitem [{\citenamefont {Schroeder}\ \emph {et~al.}(2006)\citenamefont
  {Schroeder}, \citenamefont {Martin},\ and\ \citenamefont
  {Lorensen}}]{schroeder2006visualization}%
  \BibitemOpen
  \bibfield  {author} {\bibinfo {author} {\bibfnamefont {W.}~\bibnamefont
  {Schroeder}}, \bibinfo {author} {\bibfnamefont {K.}~\bibnamefont {Martin}},\
  and\ \bibinfo {author} {\bibfnamefont {B.}~\bibnamefont {Lorensen}},\
  }\bibfield  {title} {\bibinfo {title} {The visualization toolkit, 4th edn.
  kitware},\ }\href@noop {} {\bibfield  {journal} {\bibinfo  {journal} {New
  York}\ } (\bibinfo {year} {2006})}\BibitemShut {NoStop}%
\bibitem [{\citenamefont {Scheins}(2004)}]{scheins2004tomographic}%
  \BibitemOpen
  \bibfield  {author} {\bibinfo {author} {\bibfnamefont {J.~J.}\ \bibnamefont
  {Scheins}},\ }\href@noop {} {\emph {\bibinfo {title} {Tomographic
  Reconstruction of Transverse and Longitudinal Phase Space Distribution using
  the Maximum Entropy Algorithm}}}\ (\bibinfo  {publisher} {Dt.
  Elektronen-Synchrotron DESY, MHF-SL Group},\ \bibinfo {year}
  {2004})\BibitemShut {NoStop}%
\bibitem [{\citenamefont {Wolski}\ \emph {et~al.}(2022)\citenamefont {Wolski},
  \citenamefont {Johnson}, \citenamefont {King}, \citenamefont {Militsyn},\
  and\ \citenamefont {Williams}}]{PhysRevAccelBeams.25.122803}%
  \BibitemOpen
  \bibfield  {author} {\bibinfo {author} {\bibfnamefont {A.}~\bibnamefont
  {Wolski}}, \bibinfo {author} {\bibfnamefont {M.~A.}\ \bibnamefont {Johnson}},
  \bibinfo {author} {\bibfnamefont {M.}~\bibnamefont {King}}, \bibinfo {author}
  {\bibfnamefont {B.~L.}\ \bibnamefont {Militsyn}},\ and\ \bibinfo {author}
  {\bibfnamefont {P.~H.}\ \bibnamefont {Williams}},\ }\bibfield  {title}
  {\bibinfo {title} {Transverse phase space tomography in an accelerator test
  facility using image compression and machine learning},\ }\href
  {https://doi.org/10.1103/PhysRevAccelBeams.25.122803} {\bibfield  {journal}
  {\bibinfo  {journal} {Phys. Rev. Accel. Beams}\ }\textbf {\bibinfo {volume}
  {25}},\ \bibinfo {pages} {122803} (\bibinfo {year} {2022})}\BibitemShut
  {NoStop}%
\bibitem [{\citenamefont {Roussel}\ \emph {et~al.}(2023)\citenamefont
  {Roussel}, \citenamefont {Edelen}, \citenamefont {Mayes}, \citenamefont
  {Ratner}, \citenamefont {Gonzalez-Aguilera}, \citenamefont {Kim},
  \citenamefont {Wisniewski},\ and\ \citenamefont
  {Power}}]{PhysRevLett.130.145001}%
  \BibitemOpen
  \bibfield  {author} {\bibinfo {author} {\bibfnamefont {R.}~\bibnamefont
  {Roussel}}, \bibinfo {author} {\bibfnamefont {A.}~\bibnamefont {Edelen}},
  \bibinfo {author} {\bibfnamefont {C.}~\bibnamefont {Mayes}}, \bibinfo
  {author} {\bibfnamefont {D.}~\bibnamefont {Ratner}}, \bibinfo {author}
  {\bibfnamefont {J.~P.}\ \bibnamefont {Gonzalez-Aguilera}}, \bibinfo {author}
  {\bibfnamefont {S.}~\bibnamefont {Kim}}, \bibinfo {author} {\bibfnamefont
  {E.}~\bibnamefont {Wisniewski}},\ and\ \bibinfo {author} {\bibfnamefont
  {J.}~\bibnamefont {Power}},\ }\bibfield  {title} {\bibinfo {title} {Phase
  space reconstruction from accelerator beam measurements using neural networks
  and differentiable simulations},\ }\href
  {https://doi.org/10.1103/PhysRevLett.130.145001} {\bibfield  {journal}
  {\bibinfo  {journal} {Phys. Rev. Lett.}\ }\textbf {\bibinfo {volume} {130}},\
  \bibinfo {pages} {145001} (\bibinfo {year} {2023})}\BibitemShut {NoStop}%
\end{thebibliography}%

\end{document}